\documentclass[12p,preprint]{aastex}
\usepackage{graphicx}
\DeclareMathAlphabet{\mathpzc}{OT1}{pzc}{m}{it}

\slugcomment{}

\shorttitle{}
\shortauthors{}

\newcommand{\co}{$^{12}$CO(2-1)~}
\newcommand{\tco}{$^{13}$CO(2-1)~}

\shorttitle{Molecular Clouds and H~II Regins }
\shortauthors{Azimlu M. \&  Fich M.}

\begin{document}

\title{Study of Molecular Clouds associated with H~II Regions}

\author{Mohaddesseh Azimlu \altaffilmark{1,2} \& Michel Fich\altaffilmark{2}}
\altaffiltext{1}{Department of Physics \& Astronomy, University of Western Ontario,
    London, ON, Canada, N6A 3K7}
\altaffiltext{2}{Department of Physics \& Astronomy, University of Waterloo,
    Waterloo, ON, Canada, N2L 3G1}

\begin{abstract}
The properties of molecular clouds associated with 10 H~II regions were studied using CO observations. 
 We identified 142 dense clumps within our sample and found  that our sources are divided  into two  categories: those with clumps that show a  power law size-line width relation 
 (Type I)  and those which do not show any relation (Type II). 
The clumps in the Type I sources  have larger power law indices than  found in  previous studies. 
The clumps in the Type II  sources have larger line widths than do the clumps in the Type I sources.  
The mass $M_{LTE}$ increases with $\Delta V$ for both  $^{12}$CO and  $^{13}$CO lines in Type I sources;  No  relation  was found for Type II sources.  
Type II sources show evidence (such as outflows) of current active star formation within the clump and we suggest that the lack of a size-line width relation is a sign of current active star formation.

For both types of sources no relation was found between volume density and size, but overall larger clumps have smaller volume density, indicating that smaller clumps are more evolved and have contracted to smaller sizes and higher densities.  
Massive clumps seem to have similar masses calculated by different methods but lower mass clumps have larger virial masses compared to  velocity integrated  (X factor) and LTE mass. We found no relation between mass distribution of the clumps and distance from the H~II region ionization front, but a weak decrease of the excitation temperature with increasing distance from the ionized gas. The clumps in collected shells around the H~II regions have slightly larger line widths but no  relation was found  between line width and distance from the H~II region, which probably  indicates that the internal dynamics of the clumps are not affected by the ionized gas.  Internal sources of turbulence, such as outflows and stellar winds from young proto-stars may have a more important role on the molecular gas dynamics. We suggest that large line width and larger size-line width power law indices are therefore the initial characteristics of clumps in massive star forming clouds (e.g. our Type I sources) and that some may evolve into objects similar to our Type II sources, where local ``second generation'' stars are forming and eliminating the size-line-width relation. 
\end{abstract}

\keywords{Star Formation: general --- H~II Regions}

\section{Introduction}

Massive stars are believed to form in giant molecular clouds (GMCs)  and within  clusters (Lada \& Lada 2003) which are  formed in dense, massive, turbulent clumps (Saito et~al.  2008).  Compared to isolated star formation from individual collapsing cores, clustered  star formation is more complicated   and therefore less well understood theoretically.  Massive star-forming regions  are not as common as low mass star-forming regions  and consequently, less frequent in nearby clouds. Therefore, high mass star formation has been far less studied in detail  compared to  low mass star formation. 
In  regions of high mass star formation, it is not only the physical conditions of the molecular clouds that govern the star formation. Newborn stars in these  regions also affect their original environment through various  feedback mechanisms (e.g. Krumholz et~al. 2010). Low mass stars form gently  in more quiescent regions and do not significantly  affect  their environment during their slow formation and evolution process. In contrast, high mass stars form in turbulent dense cores \citep{mckeetan03}  and evolve very quickly. They influence the environment  by strong winds, jets and  outflows which change the physical conditions such as temperature, density and turbulence of the cloud (e.g.  Gritschneder et~al. 2009). This feedback may even affect the Initial Mass Function (IMF) (e.g. Bate \& Bonnell 2005, Krumholz et~al 2010). Massive stars  also ionize the  gas  (known as H~II regions)  that expands into the surrounding cloud. All of these energetic processes near the young massive star affect further star formation in the cloud (e.g. Deharveng et~al. 2005).

We are making a detailed examination of the dense gas properties in star-forming regions  associated with massive stars in order to investigate how the gas physical parameters and consequently, star formation have been affected in these environments. Our sample consists of 10 H~II regions  selected from the Sharpless catalog of H~II regions (Sharpless 1959). 
Some of these regions have been investigated  in previous studies and it has been observed that they have complex spatial and kinematic structures (Hunter et al. 1990). Triggered star formation also has been observed at the peripheries of two sources within our sample (S104 and S212;  Deharveng et al. 2003  , 2008).  Here we study the properties of the molecular gas in higher resolution and investigate how physical parameters vary  due to the influence of the H~II region and the exciting star. 

For each source we used   
 the James Clerk Maxwell telescope (JCMT) to make   \co maps around  the H~II regions to study the clumpy structure of the gas in the associated  molecular clouds  and to locate the dense cores within the cloud. These maps can be made in poor weather conditions at the JCMT as the emission is very strong. \co  is optically  thick and cannot trace the dense gas at  the centre of the cores  where the star formation actually takes place.  Therefore, we used pointed 
observations of an optically  thinner emission line such as  \tco  in the cores to measure physical properties of the  dense gas, such as density, temperature, clump masses and velocity structure  that affect the star formation process.  In particular, the velocity structure and line widths will provide information on the dynamical  forcing by the H~II region and on the clumps'  
support mechanisms. We can also get a picture of internal dynamics  inside the molecular cloud. Hot clumps that show  evidence of outflows can also be found and identified as candidates for 
proto-stars. 

We introduce the sample and present the observational details in \S\ref{sampleobservation}. In \S\ref{results} the observational results and calculated parameters are presented. We discuss  the relationships between derived physical characteristics in \S\ref{discussion} and conclusions are summarized in \S\ref{summary}.

\section{Sample Selection and Observation }
\label{sampleobservation}

\subsection{Sample}
We have selected H~II regions with small angular size ($\leq 7'$) to be able to map the molecular gas at the edges of the ionized gas. The observed sources are listed in column one of  Table \ref{sample}.
We have selected objects in the outer Galaxy, primarily along the Perseus arm to minimize the confusion with  background sources and to have  the best estimate of the kinematic distance for  sources for  which we do not have a direct  distance determination.  Knowing the accurate source distance is crucial to estimate size and mass of the clumps detected within each cloud, therefore the sources which does not have a measured distance in literature have not been selected. S104 is the only source with almost the same Galactic orbit as the Sun, but we have a direct distance measurement for this object and it satisfies our other requirements.  All distances are given with literature references in Table \ref{sample}.

Columns two and three present the approximate  coordinates of the centre of   the mapped region in $^{12}$CO(2-1). Column four shows the distance of the  source based on the identified  exciting star from the literature; for S196 no exciting star has been detected and therefore we use the  kinematic distance for this source.  In column five we list the radial velocity of each source from the catalog of CO radial velocities toward Galactic H~II regions  by Blitz, Fich and Stark (1982).
We have selected H~II regions with small angular size to be able to map the molecular gas at the edges of the region. Column six gives the angular diameter  of optically visible ionized gas.  The angular diameter varies between one  and seven arc-minutes. Our sources lie at distances  between $\approx 1$ kpc (S175) and  $\approx7$ kpc (S212). The calculated diameters  of the H~II regions in our sample vary from  smaller than 1 pc  for S175 and  S192   to 9 pc for S104.  The calculated diameters are listed in column seven. 
 In column eight we list the identified exciting stars from the literature.

To investigate the effects of the massive exciting star and the H~II region on its environs we   compare the  molecular gas  near the H~II region with the  gas from parts of the cloud which are distant  enough that they are unlikely  to be affected. We  observed an excellent example  of a second  distinct and distant component   in the molecular cloud associated with S175  (Azimlu et al. 2009, hereafter  Paper I).  Two components in the molecular cloud,  S175A and S175B,  have previously  been identified in an IRAS  survey of  H~II regions by Chan \& Fich (1995).  Both regions have the same V$_{LSR} \approx -50$ km s$^{-1}$ and are connected by a recently observed filament of molecular gas with the same velocity.  Therefore it is reasonable to assume that S175A and S175B lie at the same distance.  We use the results from Paper I as a template to study the properties of the clouds discussed in this paper.

\subsection{Observations with the 15 m JCMT Sub-millimeter Telescope}

The observations were carried out in different stages. In the first stage from August to November 1998   we made  $7'\times7'$  \co maps of S175A, S175B, S192/S193, S196, S212 and S305.   The details of these observations  have been described in Paper I. Pointed observations in \tco on peaks of identified  dense clumps within each region were made from August 2005 to February 2006.  In this period we also made  a   $7'\times7'$ \co map of S104 with the same observational settings as in 1998, but due to the large angular size of this H~II region we missed most of the associated molecular gas in our map. Therefore, we decided to extend the S104 map to a larger area.

Further observations were made with the  new ACSIS system at the JCMT  between October 2006  and January  2009.  We extended the S104 map to a  $9'\times9'$ map and made four more  $7'\times7'$ \co maps: around S148/S149, S152, S288 and S307.  Most of the peaks identified in clumps within all sources in our sample  were also  observed in \tco during four observation missions in this period. 
In this new configuration  with ACSIS, we used a bandwidth of 250  MHz with 8190 frequency channels,  corresponding   a velocity range of about 325 km s$^{-1}$ for \co and 340 km s$^{-1}$ for \tco  with a resolution of $\sim$0.04 km s$^{-1}$.  We used the Starlink SPLAT and GAIA packages to  reduce the data, make mosaics, remove baselines and fit Gaussian functions to determine  $\Delta$V, the  FWHM of the  observed line profiles.

\section{Results}
\label{results}
 
 The physical properties   of the clumps are derived from the \co  maps and \tco  pointed observations at the peaks.
 We used the $^{12}$CO(2-1) maps to study the structure and morphology of the molecular
clouds and to identify  clumps  within them.  The cloud associated with each H~II region consists of various clumps of gas that display one or more peaks of emission (typically towards the centre of the clump).  We define any separated condensation as a distinct clump, if all of the following conditions are met: 1) the brightest peak within the region has an antenna temperature larger than five times the rms of the background noise;  2)   the drop in  antenna temperature between two adjacent bright peaks  on a straight line in between,  is  larger than the background noise; and, 3) the size of the condensation is  larger than the telescope  beam size ($21''$ or 3 pixels).   The edge of a clump is taken to be the boundary at which the integrated intensity over the line  drops to below half of the highest measured integrated intensity within that clump. An ellipse that best fits to this boundary  is used to calculate  the clump size and total integrated flux measurements.
 
 The positions  of the \co peak of identified clumps in each region and other observed parameters for each clump  are listed in Table \ref{tbl2}. Most of the clumps do not have a circular shape, therefore an equivalent  value for the radius is calculated from the area covered by  the clump   in  $^{12}$CO maps. We  define the effective radius  for each clump based on the area of the clump as  $R_{e} = \sqrt{Area/\pi}$. The corrected antenna temperatures for \co and \tco, $^{12}T_{a}^{*}$ and $^{13}T_{a}^{*}$ and the peak velocities  for each emission line, $V_{12}$ and $V_{13}$, are directly measured from spectra after baseline subtraction, and  are  listed in Table \ref{tbl2}.
 
We calculate the \tco  optical depth and hydrogen column density assuming local thermodynamic equilibrium (LTE) conditions. Detailed discussion of the methods and the equations  are  given in Paper I.  We use the  \co    peak  antenna temperature to calculate the brightness temperature and estimate the excitation temperature, $T_{ex}$, for each clump. The optical depth, $\tau$,  and $^{13}$CO column density can be derived by comparing the  $^{12}$CO(2-1) and  $^{13}$CO(2-1) brightness temperatures.  To calculate these parameters we assumed that $^{12}$CO is optically thick and  $\tau_{13CO} \ll \tau_{12CO}$. These assumptions may underestimate the \tco column density which has been considered in equilibrium conditions in section \S\ref{clumpmass}.
We  convert the  $^{13}$CO column density to hydrogen column density, $N(H)$, assuming an abundance factor of $\simeq 10^{6}$  (Pineda et al. 2008).  Derived parameters are listed in Table \ref{tbl3}.  Column two shows the excitation temperature. In columns three and four we present the FWHM of the Gaussian best fit to the $^{12}$CO and $^{13}$CO spectra. Column five presents the hydrogen column density calculated from the LTE assumption. Calculated optical depth for $^{12}$CO and $^{13}$CO are listed in columns eight and nine. 

We used three different methods to estimate the clump masses. 
The virial mass was calculated by assuming that the clumps are in virial equilibrium. We used the $^{13}$CO line widths and assumed  a spherical distribution with  density  proportional to  $r^{-2}$    (McLaren 1988). 
In Paper I, we discussed the uncertainties of this mass estimation method. For example, the line profiles are broadened  in many cases (for example strongly in S152 and S175B) due to internal dynamics and probably turbulence. As a result, the virial equilibrium assumption over-estimates the mass of the clumps and in particular is not an appropriate  mass estimation method for dynamically active regions.

We also calculated  $^{13}$CO column density under the  LTE assumption and converted it to hydrogen column density in order  to determine  clump masses. There are some uncertainties in the mass estimated this way from several factors   related to  non-LTE effects.  For example, in the LTE assumption the CO column density directly varies with $T_{ex}$.   The $^{12}$CO emission might be thermalized even at small densities, while the less abundant isotopes may be sub-thermally excited (Rohlf \& Wilson 2004).
LTE calculations assume  that  the excitation temperature is constant throughout the cloud. $T_{ex}$ is calculated from the observed $^{12}$CO line which is optically thick and only traces the envelope around the dense cores; however, inside the envelope, the core might be hotter or cooler. If the core is hotter than the envelope, the emission line might be self absorbed and  the measured antenna temperature will give too small a value of $T_{ex} $ for most of the clumps. The $M_{LTE}$ is the smallest calculated mass for most of the clumps in our sample (see also Frieswijk et.al (2007) for similar results in a dark cloud in an early stage of star formation). 

In the case of the optically thick $^{12}$CO emission we can calculate the column density from  the empirical relation $N(H_2) = X \times \int T_{mb}(^{12}CO)dv$, with  the brightness temperature, $T_{mb}= T_a^*/\eta_{mb}$. For the JCMT,  $\eta_{mb}= 0.69$ is the beam efficiency at the observed frequency range. However the ``X-factor'' is sensitive to variations in physical parameters,  such as density, cosmic ionization rate, cloud age, metallicity  and turbulence (e.g. Pineda  2008,  and references therein). The ``X-factor'' also depends on the cloud structure  and varies from region to region. A roughly constant  $X$ value is  accepted for observed  Galactic molecular clouds, and is currently estimated  at $X\simeq 1.9\pm0.2  \times 10^{20}$ cm$^{-2}$(K km s$^{-1})^{-1}$  (Strong \& Mattox 1996). 
Derived integrated column density is used to calculate the velocity integrated mass or X-factor mass. The velocity integrated mass,  $M_{int}$ and relevant volume density, are calculated for all the identified clumps and  are listed in columns six and  seven of Table \ref{tbl3}.  
In general $M_{int}$ is intermediate between $M_{vir}$, which often overestimates the mass determinations, and $M_{LTE}$, which may underestimates the mass.
Accordingly, we use $M_{int} $ as the best mass estimation for our clumps in the rest of this paper and discuss the estimated mass relations further  in \S\ref{clumpmass}. 

\section{Discussion}
\label{discussion}

In this section we discuss the physical conditions of 142 clumps identified within clouds associated with ten H~II regions. We investigate the relationships between different measured and calculated parameters  such as size, line widths, density and mass in order to understand how they have been affected in different regions. We also explore how these physical properties vary with distance from the ionized fronts of the H~II region.

\subsection{Size-Line Width (Larson) Relationship}
\label{linewidth-size}

Studies of the molecular emission  profile and line width provides us with information on the internal dynamics and turbulence of the clumps and cores. Various studies of molecular cloud clump/cores have examined the relation between line width and clump/core size (e.g., Larson 1981,  Solomon et al. 1987, Lada et al. 1991 , Caselli \& Myers 1995,  Simon et al. 2001,  Kim \& Koo  2003) and a power law, often known as the Larson relation, has been proposed to describe  this relationship:

\begin{equation}
\Delta V \propto r^{\alpha},  \hspace{1cm}   0.15 <\alpha< 0.7.
\end{equation}

\noindent This relation has been observed over a large range of clumps/cores, from smaller than 0.1~pc up to larger than 100~pc. Different power law indices have been observed within different samples. The observed relationship is  presumably affected by the cloud physical conditions, clump definition, and dynamical interaction with  associated sources such as H~II regions, newborn stars or proto-stars. 
In their survey for dense cores in L1630, Lada et~al. (1991) noticed that the existence of the Larson relation is highly dependent on clump definition. They found a weak correlation for clumps selected by  5$\sigma$ detection above the background but no correlation for clumps selected at 3$\sigma$.  Goldbaum et~al. (2011, submitted) recently used virial models to show that some GMCs' properties including the line width and size are highly dependent on the mass accretion rate and that the clouds with larger mass, radii and velocity dispersions must be older.

Kim \& Koo (2003) found a good correlation between size and line width for  both $^{13}$CO and CS observations with $\alpha=0.35$. However,  other studies show more scattered plots (e.g. Yamamoto et~al. 2006, Simon et~al. 2001) or no relation at all (e.g. Azimlu et al. 2009, Plume et al. 1997). In a study of three categories of   clumps containing massive stars,  stellar or proto-stellar identified sources, or no identified source,  Saito et~al. (2008) found a weak relation. In this study, the clumps containing  massive stars had larger $\Delta V$s.  
In a study of cloud cores associated with water masers, Plume et~al. (1997)  noted  that the size-line width relation breaks down in massive high density cores, which systematically had higher line widths.  Line widths larger than those  expected from thermal motions are thought to be due to local turbulence (Zuckerman \& Evans, 1974).  Larson  (1981) noted that regions of massive star formation such as Orion  seem to have larger $\Delta V$ and probably show no correlation with size.  Goldbaum et~al. (2011, submitted) suggested  that accretion flow is the main source of   turbulence in massive star forming GMCs, however the virial parameters remained roughly constant as the clumps evolved in their model, therefore they were able to reproduce the power law Larson relation.
Plume et. al. (1997) concluded that a  lack of the Larson relation in their data indicates that physical conditions in  very dense  cores with massive star formation are very different from  local regions of less massive star formation (the line widths may have been affected by the star formation process). They suggested that these conditions (denser and more turbulent than usually assumed) may need to be considered in studying the massive portion of the Initial Mass Function.  

This argument agrees with the conclusions of Saito et~al. (2007). In their study Saito et~al. found  clumps with no massive star to have similar line widths with cluster forming clumps classified by a previous study (Tachihara et~al. 2002) and the massive clumps observed by Casseli \& Myers (1995) all have a similar line width. These regions are all forming  intermediate mass stars, suggesting  that there is  a close relation between the characteristics of the formed stars and the line width. 
Saito et~al. also mention that, although the line width might be influenced by  feedback of young stars, extended line emission could be a part of the initial conditions of the cloud. We discuss later in section \ref{paramdist} that in our sample large line widths seem to be the initial condition of massive star forming regions.

The index $\alpha$  seems to depend on the physical conditions of the cloud and especially varies in turbulent regions (e.g. Caselli \& Myers 1995; Saito et~al. 2006). 
Star forming regions are turbulent environments and  the star formation process is believed to be governed by supersonic turbulence (e.g.  Padoan \& Nordlund 1999) driven on large scales. A turbulent  cascade then transfers the energy  to smaller scales and forms a hierarchical clumpy structure (Larson 1981). Numeric analysis of decaying turbulence in an environment with a small Mach number is consistent with the Kolmogorov law, but supersonic magneto-hydrodynamic turbulence results in a steeper velocity spectrum (Boldyrev 2002 and references therein).

We investigated the Larson relation in our sample for both \co and \tco lines. Size and line width are  directly  measured from the observations.  
Most of the previous studies has used the common Least Square (LSq) fitting method but because there are uncertainties in both $\Delta V$ and $R_e$, we prefer to calculate the slope of the fitted line with  a bisector least-squares fit (Isobe et~al. 1990).  
However we also repeated all fits with the LSq method to compare our results to previously published results (Table \ref{comparison}). 
We also calculated correlation coefficients (and significance) of the fits for the clumps associated with each of the ten H~II regions. 
In six regions a linear relationship was found in log($\Delta V$) versus log($R$) with slopes between 1.2 and 3.0 (bisector method) and between 0.44 and 1.54 (LSq). 
There are 24 fits for these six objects: six objects, two spectral lines, and two fitting techniques.
On average the slopes found in the bisector were 1.4 sigma greater than the slopes found from the LSq method.
in most of the fits the $^{12}$CO and $^{13}$CO slopes were the same to within the uncertainties.
In all six objects a modest correlation was found (typical correlation coefficients of 0.57) with 1.5 to 3 sigma levels of significance.
The slopes and uncertainties for both $\Delta V_{12}$  and $\Delta V_{13}$  are listed in Table \ref{slopes} for these six regions - those with a linear relationship. 

In the other four regions no size-line width relation was found for any fitting technique with either set of CO data. 
In all four objects the best fit slopes were close to zero and the uncertainties in the slopes were large and the correlation coefficients are close to zero - with the exception of S175B which had a coefficient of -0.59 at the 1.8 sigma level in the $^{12}$CO data.

We therefore  divided  our sources into two categories: the ``Type~I"  sources that  show a size-line width relation and  ``Type~II'' sources with a weak or no Larson relation. The clumps within Type~II  sources  have broader  lines  in general and have a scattered size-line width plot. 
Figure \ref{delvrnt} shows the plots  of $\Delta V_{12}$ and $\Delta V_{13}$ vs $R_e$ and the derived $\alpha$s (bisection method) for both \co (black dots) and \tco (red dots) lines for Type~I sources.  S288 has only 4 identified clumps and we therefore excluded this region.  Figure \ref{delvrt} shows the same plot as Figure \ref{delvrnt} for Type~II sources. No power-law lines are shown on these plots as no relation could be obtained for these objects. 

The LSq slopes we have calculated for Type I regions are larger than the previous reported values (e.g. Larson 1981,  Solomon et al. 1987, Lada et al. 1991, Caselli \& Myers 1995,  Simon et al. 2001,  Kim \& Koo  2003, Saito et~al. 2006, 2007, 2008).  
This might be due to the small samples - typically a dozen clumps in each region - which also leads to relatively large uncertainties in the calculated slopes.
All of the previous studies have faced the same difficulty and solve the problem by combining the data from all objects; In some studies data from different spectral lines is combined together for this analysis.

 Considering all six Type I sources together, we derived a slope of   $\alpha=0.76\pm0.05$ for $\Delta V_{12}$  vs. $R_e$ and $\alpha=0.71\pm0.05$ for $\Delta V_{13}$ vs $R_e$ with correlation coefficients of 0.70 and 0.62 respectively. 
Using the LSq technique we get a slope of   $\alpha=0.51\pm0.06$ for $\Delta V_{12}$  vs. $R_e$ and $\alpha=0.47\pm0.07$ for $\Delta V_{13}$.
The results are listed in Table \ref{comparison}. We have also checked the results and generated the same statistical information for three recent similar studies and show these results in this table.  
The second column gives the slope and uncertainty while the third column shows the scatter around the best fit line.  The correlation coefficients for each data set is shown in the last column.
All of the data-sets show similar scatter. In all but the Type II sources  case we have a  correlation that is significant.  In the Type II sources  case the correlation coefficient is small and the probability of getting this just by random is large. Our Type I sources show the steepest slope, only marginally more (1/2 sigma) than more than found by Saito et al. (2006).

Type II sources are clearly and significantly different.  
The Type II  sources  have larger line widths in both $^{12}$CO and $^{13}$CO (Figure \ref{delvrt}). 
Such large line widths may originate from the initial conditions of the clump or may be caused by  energy inputs such as proto-stellar  outflows, radiation pressure from massive stars, strong stellar winds, internal rotation and infall to proto-stars within the clumps.  
All of the Type II regions show signatures of active star formation. 
S104 and S212 are good samples of triggered star formation by ``Collect and Collapse'' process (Deharveng et~al. 2003 and 2008).
We have detected strong signatures of an outflow within S175B (Azimlu, et~al., in preparation) which has the largest line widths in our sample.  
S152 is the other source in our sample with similar large line widths. This region has very active  star formation  and contains a dense  stellar cluster (Chen et~al. 2009, also detected in 2MASS data). 

To provide enough gravitational  energy to bind the clumps with larger internal velocity dispersion,  much higher densities  are required than for the clumps with smaller line widths (Saito et~al., 2006). 
High gas density ($n >10^{5}~$cm$^{-3}$) is an essential factor in  the formation of   rich embedded clusters  such as the one detected in S152  (Lada et~al. 1997). 
In addition, these clumps  must be gravitationally bound in order to survive longer than the star formation time scale of $\sim 10^6$ yr. In a massive star-forming model, McKee \& Tan (2003)  suggest that the mass accretion rate in a core embedded in a dense clump depends on the turbulent motion of the core and the surface density of the clump. Therefore, to form a dense core that can produce a massive star,  the clump requires both a high gas density and large internal motion while it is  also  gravitationally bound. 
External pressure may have an important role in providing additional force to keep the clump stable during the star formation process (Bonnor-Ebert model; Bonnor 1956, Ebert 1955). 
We have detected such dense, hot clumps within our sample (for example S104-C4, S305-C7, and S307-C4). We already have deep CFHT near IR observation for five of our sources. Data is partially reduced and we will use that to study the luminosity function and mass distribution of the young stars embedded within the clumps in our sample to determine the influences of gas properties on star formation process. 

\subsection{Size-Density Relation}

We  investigated the relation between column density and size for Type I  and Type II   sources as one may expect larger column density for larger clumps. 
In  Figure \ref{ncolr} we plot N(H) versus clump effective radius for both  Type I  (left panel) and Type II  (right panel) sources. We found statistically significant no relation between the column density and size for our sources.  We also investigated  the relation  between  velocity integrated  volume density and size. Results are presented in Figure \ref{nintrnt}.  There is a  weak  relation  for some individual sources such as S175A and S175B but, in total,  the density decreases as size  increases.  The dashed line with a slope of -2.5  shows the limit below  which we have not found  any clump.  Simon et~al. (2001)   found different power law indices  for their sample of four molecular cloud complexes varying  between -0.73 and -0.88, slightly smaller than the power law factor of -1.24 found by Kim \& Koo for molecular clouds associated with ultra compact H~II regions.  These results may be due to the fact that the smaller clumps are probably  more evolved. The gas is more collapsed  to the centre and therefore these clumps have smaller size with larger volume densities.  This also  explains the lack of  a relation between size and column density. The smaller clumps have higher volume density and integrating  density through the clump centre may result in a small or large column density dependent  on the initial conditions and the nature of the clumps. 

Sources within our sample are located at different distances. We have detected the smallest  clumps in S175A which is  the closest source to us at a distance of $\approx 1$ kpc. If this source was at  five times the distance (approximately the distance of S305),  clumps C1-C4 and  C11-C13 would not be resolved and we would measure  smaller average volume density for these groups of clumps as  individual objects. Thus, our plot is likely to  be  affected  by  spatial resolution  effects. 

Saito et~al. (2006) suggest that the mass and density must be larger for turbulent clumps (they define a core as turbulent if  $\Delta V > 1.2$ km s$^{-1}$) compared to non-turbulent   ones for a given size to bind the turbulent clumps gravitationally. 
For a similar density distribution, they found that turbulent clumps (those  with larger line widths) have densities twice the non-turbulent   ones. 
We do not see a difference in general between mass or column density of our Type II (with larger line widths) and Type I sources. For a detailed  discussion on  different masses and equilibrium conditions of the clumps see \S \ref{clumpmass}.

\subsection{LTE Mass and Column Density-Line Width Relation}

 We  find  a weak relation between N(H) and $\Delta V$ (Figure \ref{ncoldelv12} and \ref{ncoldelv13}).  The column density increases with $\Delta V$  for both $^{12}$CO and $^{13}$CO lines. The dashed  lines present the bisector least-squares  fits with  slopes of 2.3$\pm0.25 $ and 2.1$\pm0.35$ for  $^{12}$CO lines in Type I and Type II sources  respectively but the correlation coefficient is poor (0.53 and 0.42 respectively).  The slopes are  similar  (2.3$\pm$0.23 and 2.4$\pm$0.38) for  $\Delta V_{13}$  for both  Type I and Type II sources with similar correlations (0.51 and 0.47 respectively). 

$M_{vir}$ and $M_{int}$ both depend on the emission line  profile and are expected  to increase for Type II  regions with larger line widths. The LTE mass calculation is independent of the cloud dynamics and the emission line profiles.  In Figures \ref{mltedelv12} and  \ref{mltedelv13} we show how $M_{LTE}$ varies with $\Delta V $ for both $^{12}$CO and $^{13}$CO emission lines.  For Type I  sources $M_{LTE}$  increases with $\Delta V$ for both $^{12}$CO and $^{13}$CO with least-square fit slopes of $4.2\pm0.28$ and   $4.2\pm0.24$, but no relation is found  for Type II  sources. The mass range is not very different for Type I   and Type II clumps ($1.1 - 350 $ vs $0.5 - 500$ M$_\odot$). The LTE mass is calculated by integrating the  LTE column density  over the area of each clump assuming that it is a sphere with radius of $R_e$.  $M_{LTE}$ is proportional to  $R_{e}^{2}$; therefore, the dependance of mass calculation on size  might be the cause  of  the relation between $M_{LTE}$ and $\Delta V$ in Type I  sources where the size-line width relation was found.

\subsection{Equilibrium State of the Clumps}
\label{clumpmass}
The equilibrium state of a clump can be determined  by comparing the kinetic and  gravitational energy density which can be obtained by measuring the ratio of  the virial and the LTE mass (Bertoldi \& McKee 1992).
The virial mass is calculated as $M_{vir}(M_{\odot})=126\times R_e$(pc) $(\Delta V)^2$(km s$^{-1})$ which assumes a spherical distribution with  density  proportional to  r$^{-2}$ \citep{McLaren88}. Column density is generally higher at the centre of clumps and drops rapidly by distance from the centre. Assuming a simple constant density will then overestimate the mass. However it is not very easy to fit a unique density profile to all clumps especially where there are more than one cores.   Virial equilibrium also assumes  that the gravitational  potential energy is in balance with internal kinetic energy.  $M_{vir}$ increases with $(\Delta V)^2$; integrated mass, $M_{int} $, is also line-profile dependent and  varies linearly with $\Delta V$, while $M_{LTE} $ is independent of line width and of the internal dynamics of the cloud.  We plot $M_{vir}/M_{LTE}$ versus $M_{LTE}$ in Figure \ref{mvirltelog} to investigate whether clumps are in virial equilibrium.  
Resolving the clumps within clouds is highly dependent on the distance to the cloud.
To decrease  the effect of  resolution  we have selected only objects at distances between 3 and 7 kpc.  Closer  sources are represented by crosses in the plot.  Most of our clumps have larger virial masses and are far above the $M_{vir} = M_{LTE}$  line, indicating  that they are probably  not  gravitationally stable.  
An external pressure is required to keep the clumps  with $M_{vir} > M_{LTE}$  bound. 
The mean pressure of the ISM in general  ($P/k \approx 3 \times 10^4  K  cm^{-3}, k$ is Boltzmann constant, Boulares \& Cox, 1990) is sufficient  to bind clumps with $M_{vir}/M_{LTE} < 3$. Many of the clumps in our sample have $M_{vir}/M_{LTE}$ between 3 and 10 where larger pressures are required, typically $10^5$ to $10^6 K cm^{-3}$.  However all of our  clumps are near HII regions, warmer areas where the pressure is expected to be larger than the mean ISM pressure.  Even larger  external pressures, greater than  $P/k \approx  10^7  K  cm^{-3}$ is required to stabilize the small number of clumps with the largest linewidths and with  $M_{vir}/M_{LTE} >10$.  Perhaps these clouds are indeed not stable.

{Saito et~al. (2007) suggest that to bind the turbulent clumps (clumps with larger line widths) they must have larger masses, therefore for similar size clumps the turbulent sources must have larger densities. 
In  our sample both Type I  and Type II  sources (which have generally larger line widths) have similar clump masses and we do not see an excess of density for either  type. 

Figure \ref{mvirltelog} shows that  more massive clumps tend to be closer to virial equilibrium. 
The same pattern has been observed by Yamamoto et~al.  (2003, 2006).  
Figure \ref{mintlte} shows the plot of velocity integrated mass, $M_{int}$, versus $M_{LTE}$. 
For most of the clumps, $M_{int}$ is higher than the  $M_{LTE}$ but mostly below  the line of the  $M_{int}= 10 M_{LTE}$.  $M_{int}/M_{LTE}$ is much larger  for low mass clumps and like  $M_{vir}$, $M_{int}$ tends to be equal to $M_{LTE}$ for higher masses; however, the difference is smaller compared to $M_{vir}$ versus $M_{LTE}$.

In Figure \ref{mvirint} we compare   $M_{vir}$ versus $M_{int}$. Similar to  $M_{LTE}$, clumps have larger  $M_{vir}$  for low mass clumps.  At approximately M= 100 M$_\odot$,   $M_{vir}\simeq M_{int}$, while for clumps larger than 100 M$_\odot$,  $M_{vir} <  M_{int}$. 
The higher mass clumps in our plot lie within more distant sources where we cannot resolve smaller structures. This  suggests that probably the average density and temperature of larger clumps results in a larger Jeans length while smaller dense structures resolved  as clumps have locally higher density and therefore smaller Jeans length.   Analysis of cloud and clump equilibria in a study of four molecular clouds at different distances shows that, while the whole cloud is gravitationally bound, the majority of the clumps within them are not (Simon et~al.  2001). We found that the equilibrium appears for clumps larger than $\sim 100 M_\odot$.

\subsection{Temperature-Line Width Relation}

In Figures  \ref{texdelv12} and \ref{texdelv13} we investigate the relation between the excitation temperature and  line widths. Part of the profile line broadening is due to the thermal velocity dispersion of the molecules and  the thermal line width increases with temperature. The observed $\Delta V$s are more broadened than calculated thermal line widths due to internal dynamics and turbulence within the clouds. We found no relation between $T_{ex}$ and $\Delta V$ for either $^{12}$CO and $^{13}$CO emission lines which indicates that the emission line profile is dominated by the internal dynamics of the clumps.

\subsection{Effects of Distance from H~II Region on Physical Parameters}
\label{paramdist}
We are studying how H~II regions and their exciting stars affect the physical conditions of their associated clouds. One might expect more influence on clumps that are closer to the H~II region.  To investigate these effects, we examine how the environmental parameters vary with distance from the edges of the ionized gas. Most of the H~II regions in our sample are not perfect spheres but we try to fit the best circle to the  visible edges of the ionized gas. To determine the borders of the ionized gas we use Digital Sky Survey (DSS) images in the red filter in which the hydrogen ionized gas is bright with sharp edges. 
Our sources have different angular sizes, different physical sizes and lie at different distances. They might also be at different evolutionary stages. We need to re-scale the distances to a common distance to be able to compare them.  A normalized distance for each clump is defined as  the distance of the clump from the edge of  H~II region divided  by the   H~II region radius.

\subsubsection{Temperature-Distance relation}

In Figure \ref{texndist} we investigate the variation of excitation temperature, $T_{ex}$, with  distance from the H~II region. We expect clumps to be warmed up by the radiation from the exciting  star and to show a trend of decreasing  temperature with  increasing distance from the H~II region.  We can see  that the clump temperature decreases with distance  from the ionized gas around  some sources such as S175A and S152. We observe a scattered relation for S305 and a weak relation (for clumps at distances larger than 0.1R) for S104. 
We cannot see the same relation for other sources perhaps due to internal heating sources such as proto-stars within the clumps. C4 and C8 in S307 (noted in Figure \ref{texndist}, left panel) are good examples of such compact clumps with high density and temperature. 
 
On the other hand, we are measuring the projected distance of the clumps which in general  will  be smaller than the true  value. We check the effect of distance projection by  numerically simulating  randomly distributed  clumps  with heating from an external source.   
The luminosity of the   heating source, the power-law index of decrease in central heating with distance, and the heating of the clump from the diffuse  background are the input parameters of the simulation. We also set the number of clumps and the radial distribution of the clumps' positions. The output is  temperature versus the  projected distance from the heating source.

Figure \ref{tsim} shows plots of some selected simulations for different initial parameters. Setting the input values  as the  observed parameters of our sample and assuming a $R^{-\alpha}$ decreasing power law for luminosity, (Figure \ref{tsim}) we  see some scattering on the plot for small distances, but a decrease in temperature as distance from the heating source increases.   Distant hot clumps cannot be  warmed by the central heating source. Such distant  but hot clumps like S307-C3 and S307-C4 noted in Figure  \ref{texndist} are probably being warmed by an internal  source such as a proto-star. 
We see more scatter  in the simulated plots for smaller projected distances. Most of our mapped regions ($7'\times7'$) are small compared to the size of the H~II region;  consequently we did not map the molecular gas at large  distance  from the  edges of H~II regions with larger angular size (e.g. S104, S212 and S148/S149). 
The scattering of $T_{ex}$ versus normalized distance for the clumps within these sources matches with the objects at the smaller distances of the simulated plots. 
S175B is the only source in  which all of the clumps are  too distant  to have been influenced by the H~II region.

\subsubsection{Line Width-Distance Relation }
\label{lw-dist}
We study  the effect of the expanding ionized gas on the internal dynamics of the molecular gas by investigating the variation of line widths with distance from the H~II region. In Figure \ref{delvndist} we plot $\Delta V$ versus normalized distance for $^{12}$CO (left panel) and $^{13}$CO (right panel).  
We did not  find a relation for $\Delta V_{12}$ or $\Delta V_{13}$ versus distance for any of Type I  or Type II  sources. The line-width is very scattered at different distances even for S175A and S152 which show a decrease of  $T_{ex}$  with distance from the ionized gas.  The expansion of the ionized gas may cause  the  molecular gas in shells around  the H~II region  to have slightly larger line widths  but we do not see a significant distance effect on the internal motions  of the clumps  beyond the collected shells. 
Physical conditions of individual clumps and other internal sources of turbulence or dynamics such as proto-stellar outflows,  infall and rotation may  have a more important effect  on  line profiles.

The important  point is that, as discussed in section \ref{linewidth-size},  we have seen larger line widths and larger Larson power law indices in our sample for Type I sources. But  if the line profiles are not much affected by the H~II region and the  exciting star, then the observed large line widths are more likely to be   the initial characteristics of the clouds which have already formed at least one massive star.

\section{Summary and Conclusions}
\label{summary}

We have studied  the physical properties  of  molecular clouds associated with a sample of ten  H~II regions. We mapped eleven  $7'\times7'$ areas in $^{12}$CO(2-1) at the peripheries of the ionized gas and extended one of these maps (around S104 H~II region, the largest region) to $9'\times9'$ in order to study the characteristics  of  the molecular gas well beyond the edges of the H~II region.  
We investigated the clumpy structure of the clouds and identified 142 distinct clumps within eleven mapped regions.   We also  made pointed observations in \tco at the position of the  brightest $^{12}$CO peak within each clump for 117 clumps. We used these observations to measure and calculate the physical characteristics of the clouds. We summarize our findings below:

\noindent We investigated size-line width relation for our sources using both  $^{12}$CO and   $^{13}$CO emission lines and calculated effective radius. Our sources are divided into two   categories:  those which  show a power law relation and those which do not show any   relation.  We labeled the first category of six regions  as Type I and the other four sources with no relation as Type II.
Type II sources have larger line widths in general and they are active star forming regions. 
 
\noindent The power law indices derived for size-line width relation in  Type I sources are somewhat larger than found in previous studies, but they do not appear to be affected by the exciting star or the ionized gas.  We  conclude that larger line widths and consequently larger indices are more likely  the initial conditions of the massive star forming molecular clouds.

\noindent  Clumps with larger column densities trap more internal radiation and are expected to be hotter. We investigated the relation between column density and temperature and found that the temperature increases with column density for both types of sources.

\noindent Dense and hot clumps have been detected in our sample. Such clumps are more stable against fragmentation and could be candidates for formation of  massive stars in future.

\noindent No relation was found between column density and size of the clump but,  for both Type I  and Type II  sources,   the volume density decreases with size; the larger clumps have smaller densities.  The smaller clumps might be  more evolved,  contracting to smaller size and  higher densities. 

\noindent We investigated the relationship between LTE column density and line width. We found that the column density increases with both  $\Delta V_{12}$ and  $\Delta V_{13}$ for both  Type I and Type II regions but the relation is very scattered.

\noindent We estimated the mass of each clump in three different ways: the virial mass using $\Delta V_{13}$ determined from optically thinner $^{13}$CO lines ($M_{vir} \propto (\Delta V_{13})^2$), $^{12}$CO velocity integrated mass or X factor mass ($M_{int} \propto \Delta V_{12}$), and LTE mass (mass estimate independent of line width) using both $^{12}$CO and $^{13}$CO  lines. 
$M_{vir}$ is larger than  $M_{LTE}$ for small clumps but tends to equal values for large masses. Low mass clumps  also have  larger $M_{vir}$  than $M_{int}$ but $M_{vir}$ becomes approximately equal to $M_{int}$ at M=100 M$_\odot$ and is smaller than $M_{int }$ for larger masses. 
We conclude that the larger clouds are gravitationally bound but the fragmented smaller clumps within them are not.

\noindent We investigated how   $M_{LTE}$ varies with $\Delta V$ for both Type II  and Type I  sources. While we see that $M_{LTE}$ increases with  both $\Delta V_{12}$ and $\Delta V_{13}$ for Type I  sources,  no relation was found for Type II  regions.

\noindent No relation was found between the excitation temperature and the line widths. This suggests that the line widths  for both $^{12}$CO  and $^{13}$CO  are  determined by the internal dynamics of the clump rather than the thermal velocity dispersion of the molecules.

\noindent Excitation temperature decreases with distance from the edges of the ionized gas for some clouds. However, the measured decrease is not significant because of the projected distance effect, especially for smaller distances.

\noindent The effect of the H~II regions  on the internal dynamics of the clumps was investigated.  The line widths for the clumps within collected shells around H~II regions are slightly larger, but no relation was found between  either $\Delta V_{12}$ or  $\Delta V_{13}$ and normalized distance from the H~II region.  The expansion of the ionized gas affects the internal dynamics of the collected mass but these effects do not go beyond the shells. The plots of $\Delta V$  versus normalized distance are very scattered for both Type I  and  Type II  regions,  even for those sources in which  temperature decreases with distance.  If the internal dynamics of the cloud are not much affected by the exciting star and the expanding ionized gas, we  conclude that larger line widths are initial characteristics of the observed molecular clouds which already have formed massive stars.

\clearpage

\begin{center}
{\bf Acknowledgment}\\
\end{center}

The JCMT observatory staff, especially Ming Zhu, Gerald Schieven and Jan Wouterloot are thanked for their support during the observations and their helpful assistance in data reduction. We thank Pauline Barmby for proof-reading  the paper and helpful comments. We also thank the  anonymous referees for detailed  comments and suggestions that helped to improve this work.
The James Clerk Maxwell Telescope is operated by The Joint Astronomy Centre on behalf of the Science and Technology Facilities Council of the United Kingdom, the Netherlands Organization for Scientific Research, and the National Research Council of Canada.
This research was supported with funding to M.F. from the Natural Sciences and Engineering Research Council of Canada. 

{}

 \begin{figure}
   \centering
   \includegraphics[width=14cm]{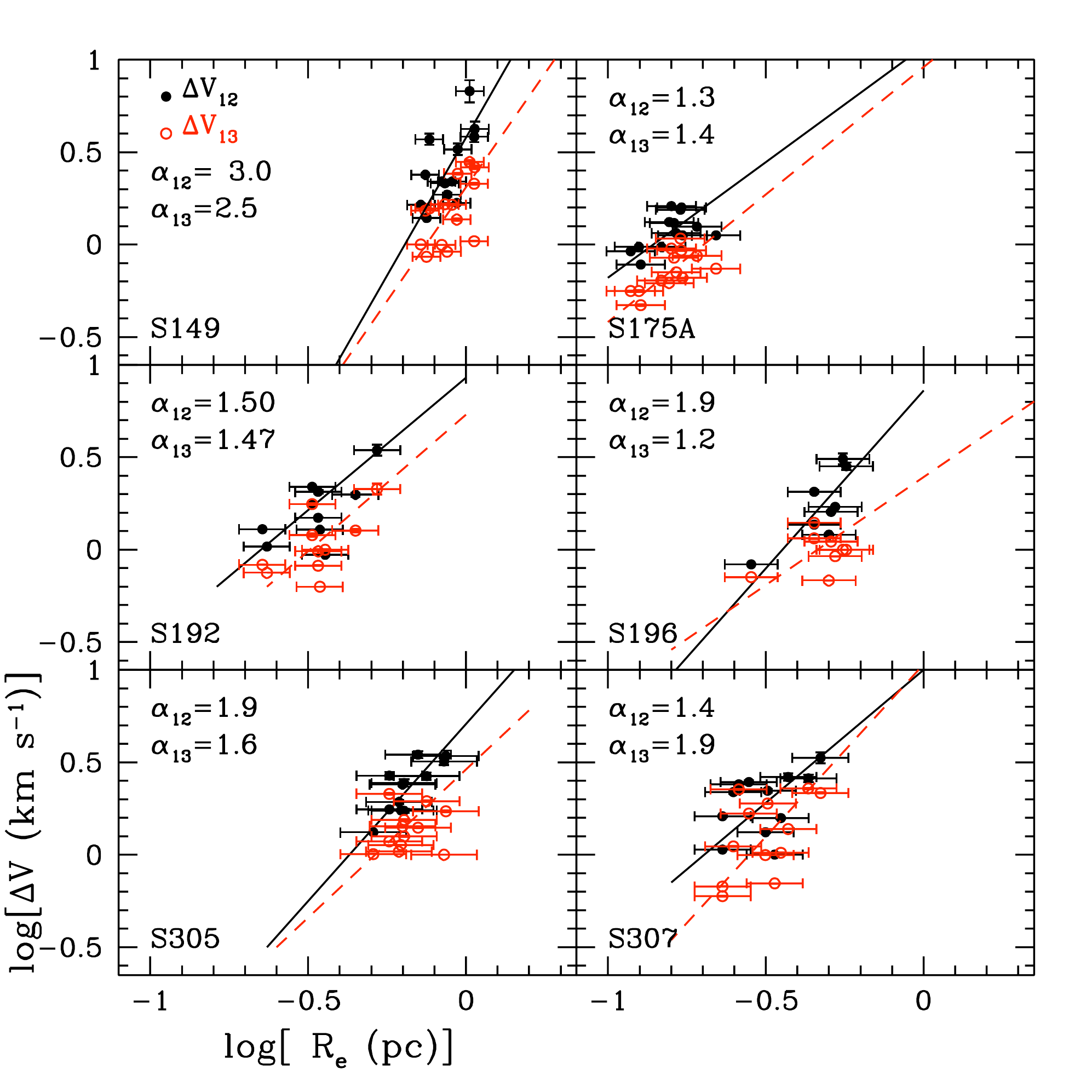} 
   \caption[Correlation between the \co  and \tco line widths and the effective radius of the clumps in Type I sources]{ Correlation between the \co (black) and \tco (red) line widths  and the effective radius of the clumps  for Type I sources. The  solid line is the  bisector least-square fit for \co and dashed line shows the bisector least-squares fit for \tco lines. The slope derived for each line is shown in  top left  corner of each panel.

   \label{delvrnt}}
\end{figure}

\begin{figure}
   \centering
   \includegraphics[width=14cm]{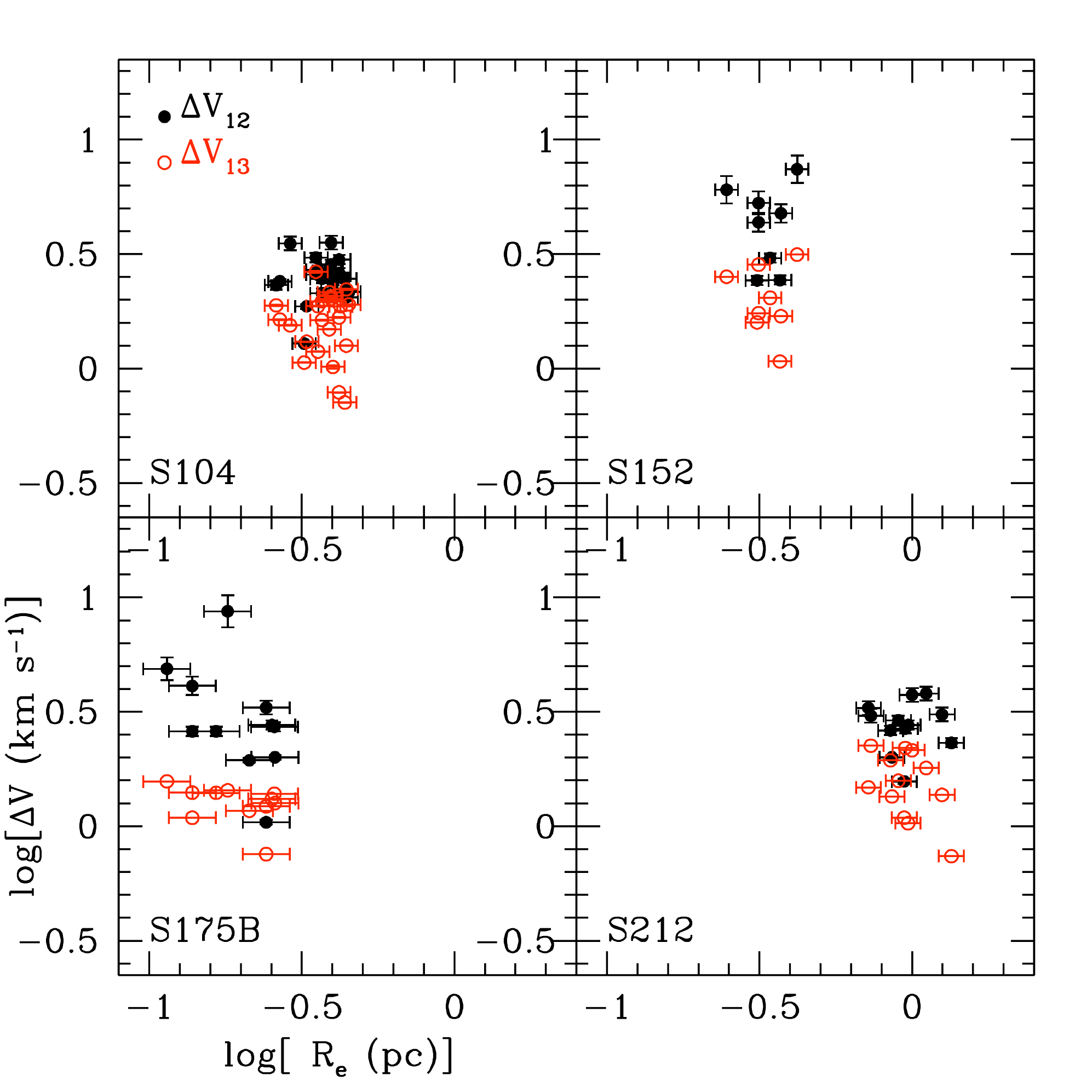} 
   \caption[Correlation between the \co and  \tco line widths and the effective radius of the clumps in Type II sources]{Same plot as Figure \ref{delvrnt} for Type II sources.      \label{delvrt}}
\end{figure}

\begin{figure}
   \centering
   \includegraphics[width=7.5cm]{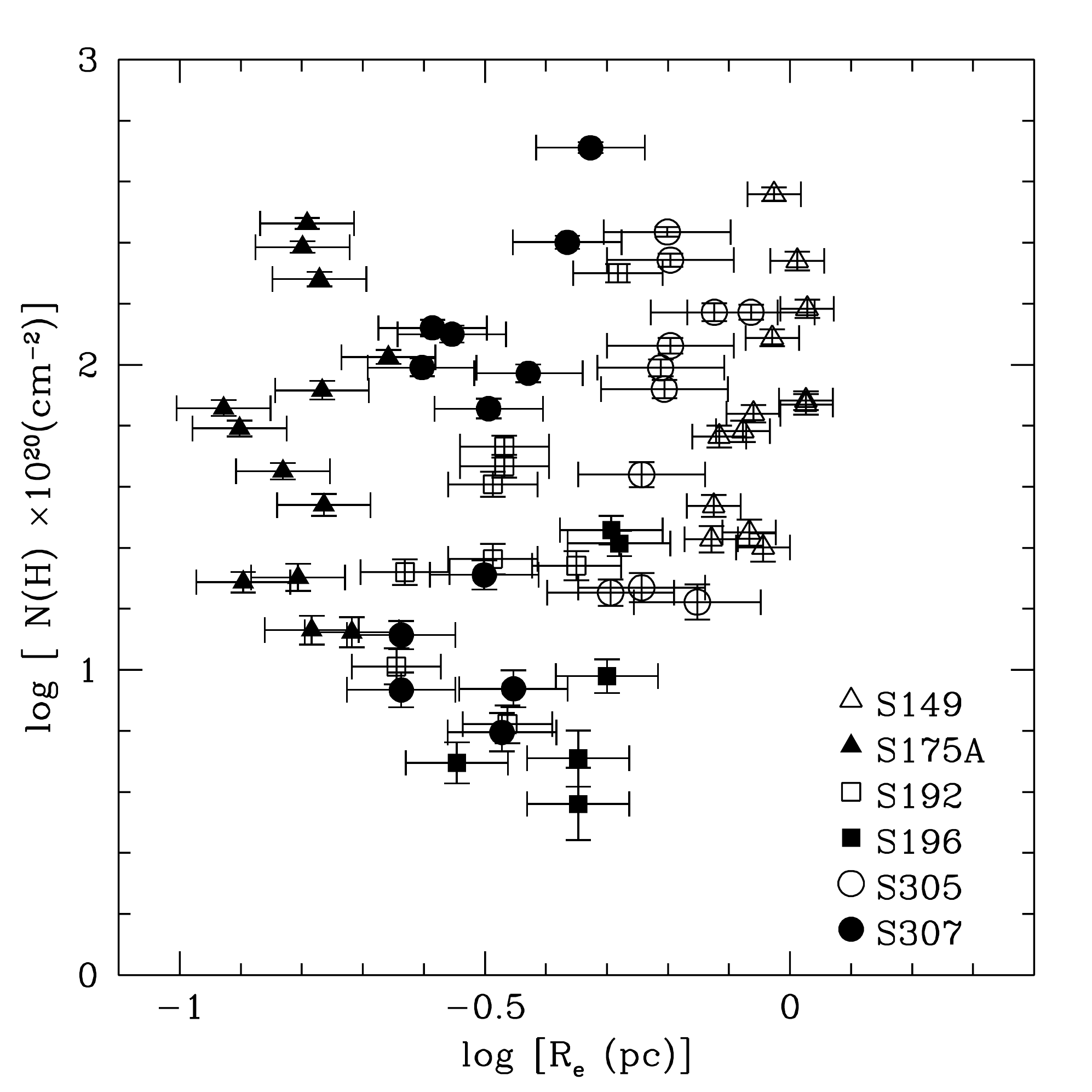}
   \includegraphics[width=7.5cm]{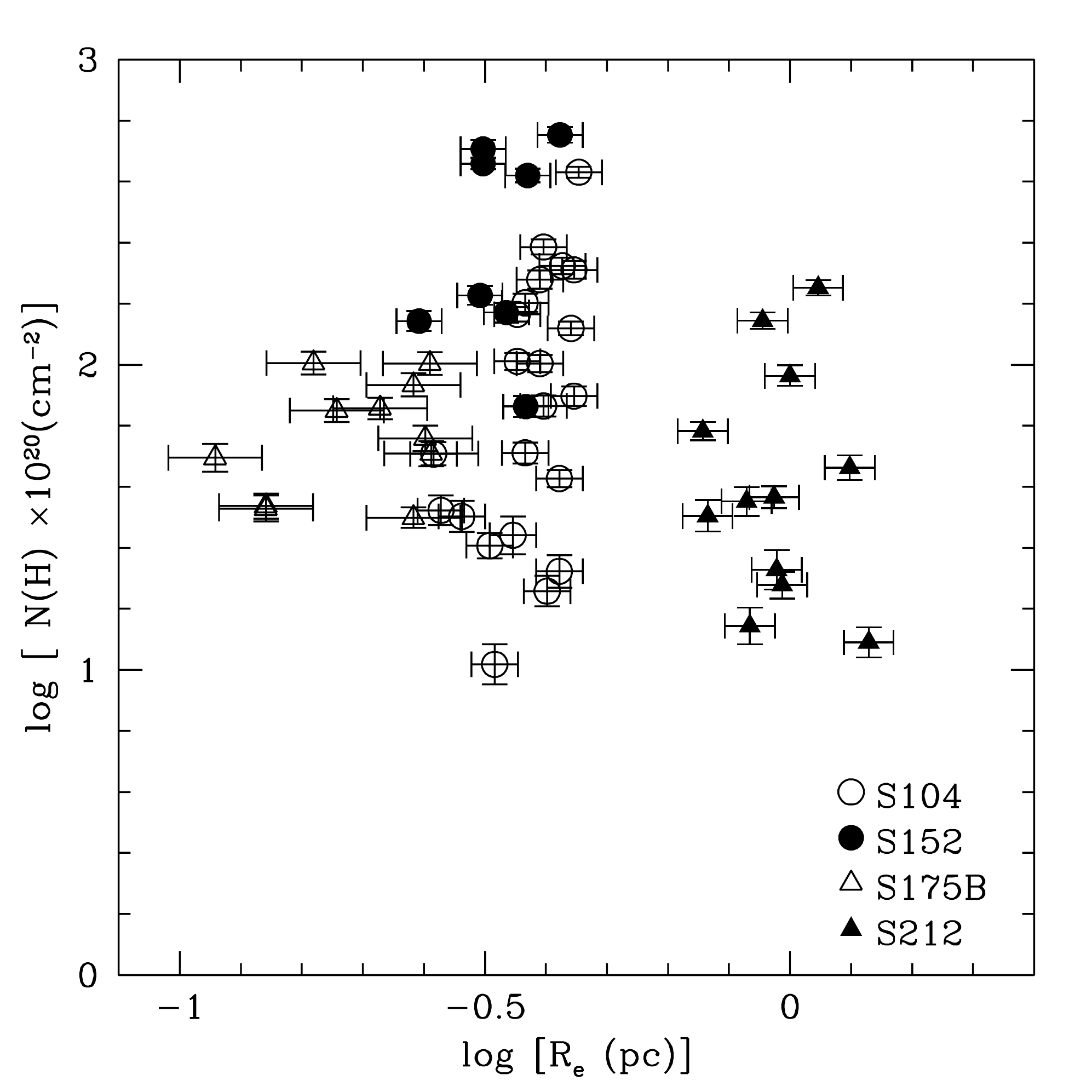}    
   \caption[LTE column density vs. effective radius]{LTE column density vs. effective radius for Type I  (left)  and Type II  (right)  sources. 
   }
   \label{ncolr}
\end{figure}

\begin{figure}
   \centering
   \includegraphics[width=7.5cm]{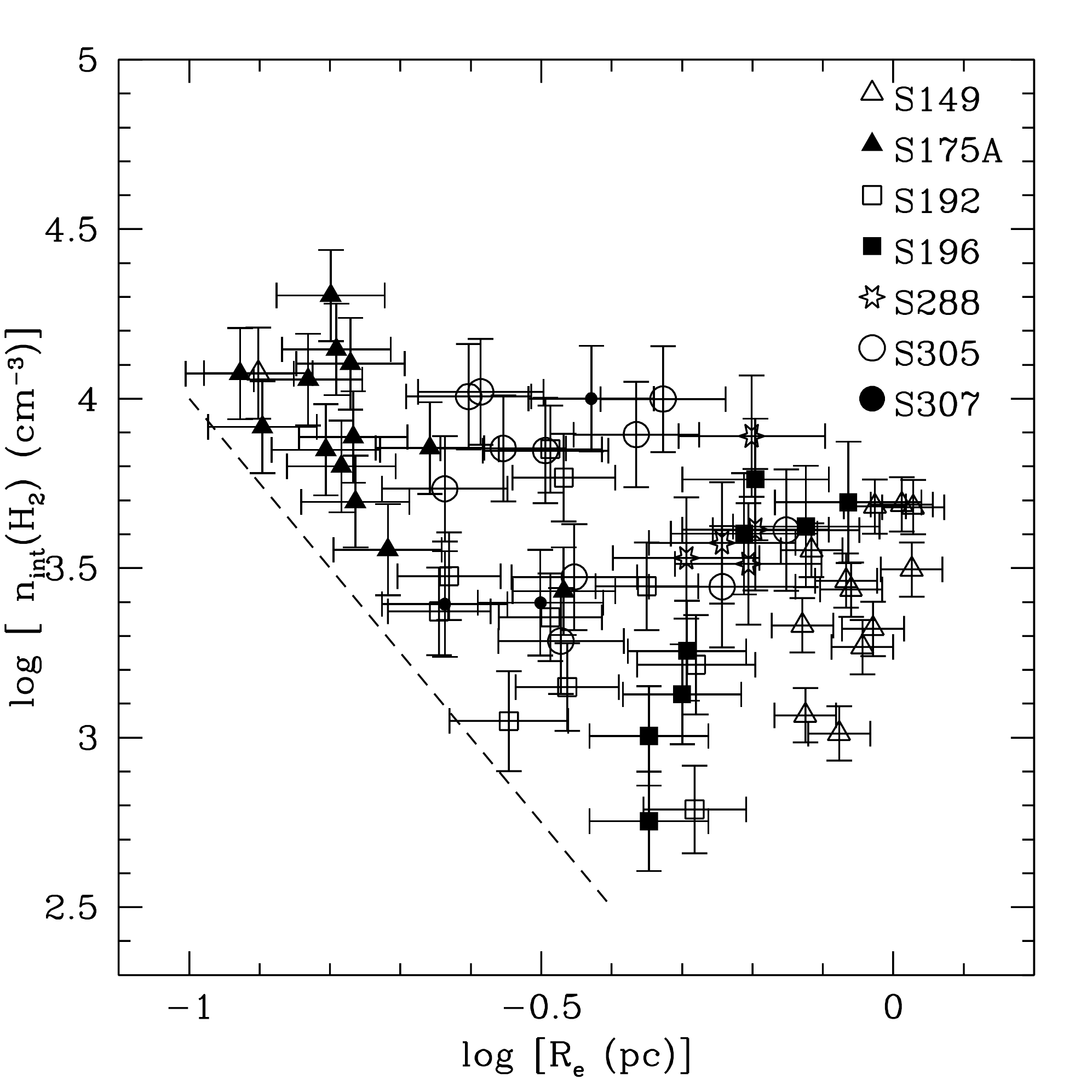}
    \includegraphics[width=7.5cm]{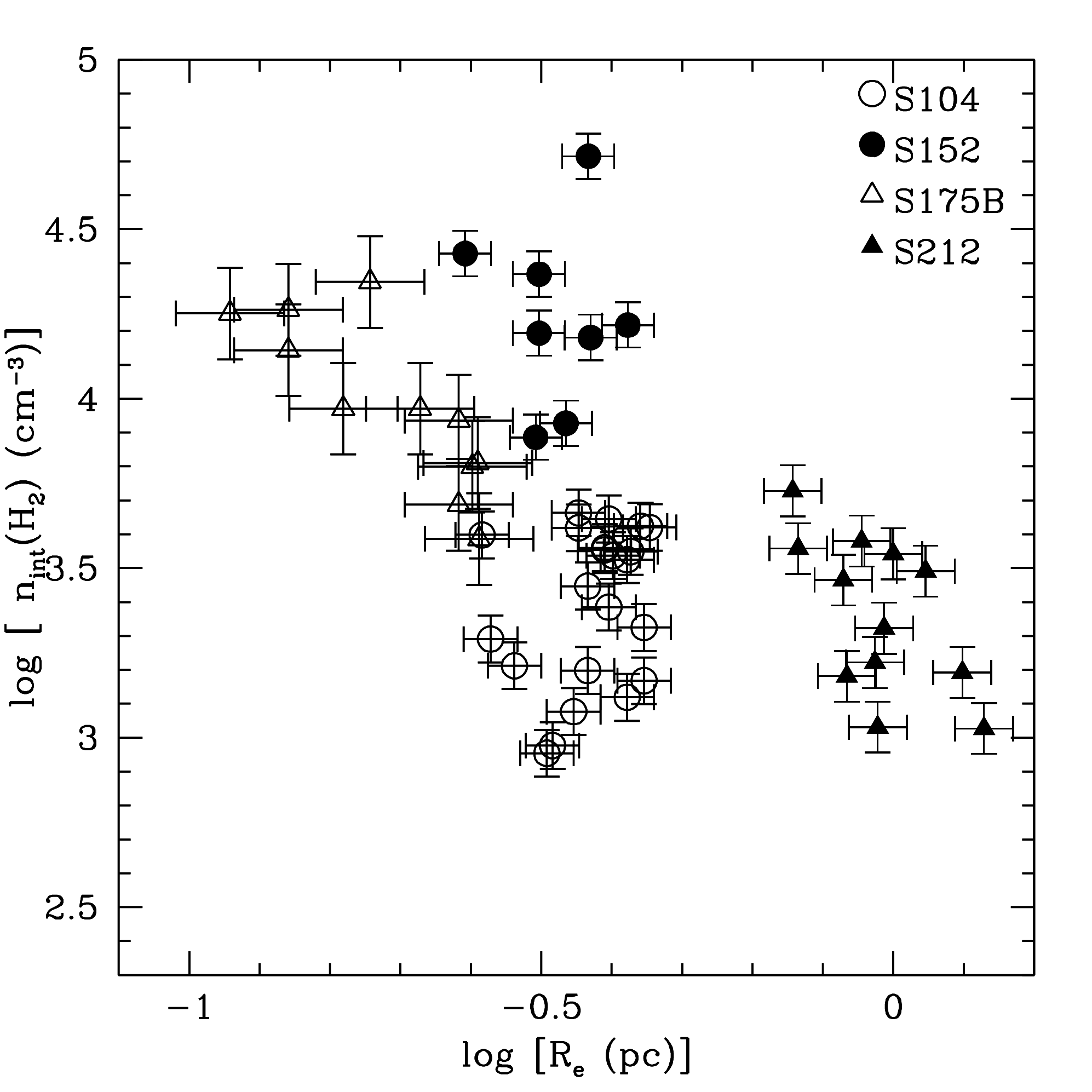}
    
   \caption{ Velocity integrated volume density  vs. effective radius for Type I  (left) and Type~II (right) sources. The dashed  line shows an approximate limit below   which no clump has been found in our data ($n_{int}=-2.5R_e+1.5$).  No strong correlation is found between $n_{int}$ and $R_e$ but  overall  the volume  density is smaller  for larger clumps. 
   }
   \label{nintrnt}
\end{figure}

\begin{figure}
   \centering
   \includegraphics[width=7.5cm]{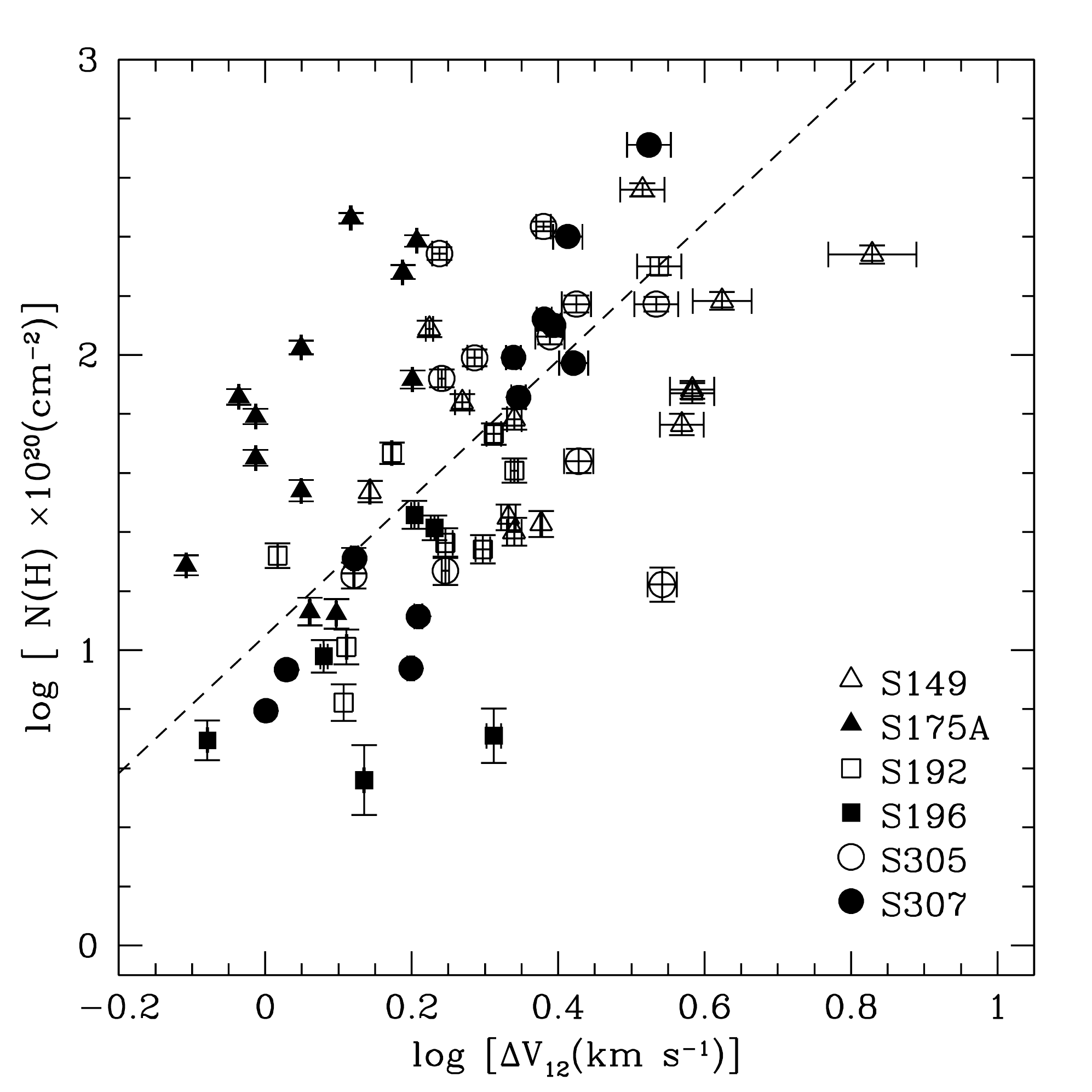}
   \includegraphics[width=7.5cm]{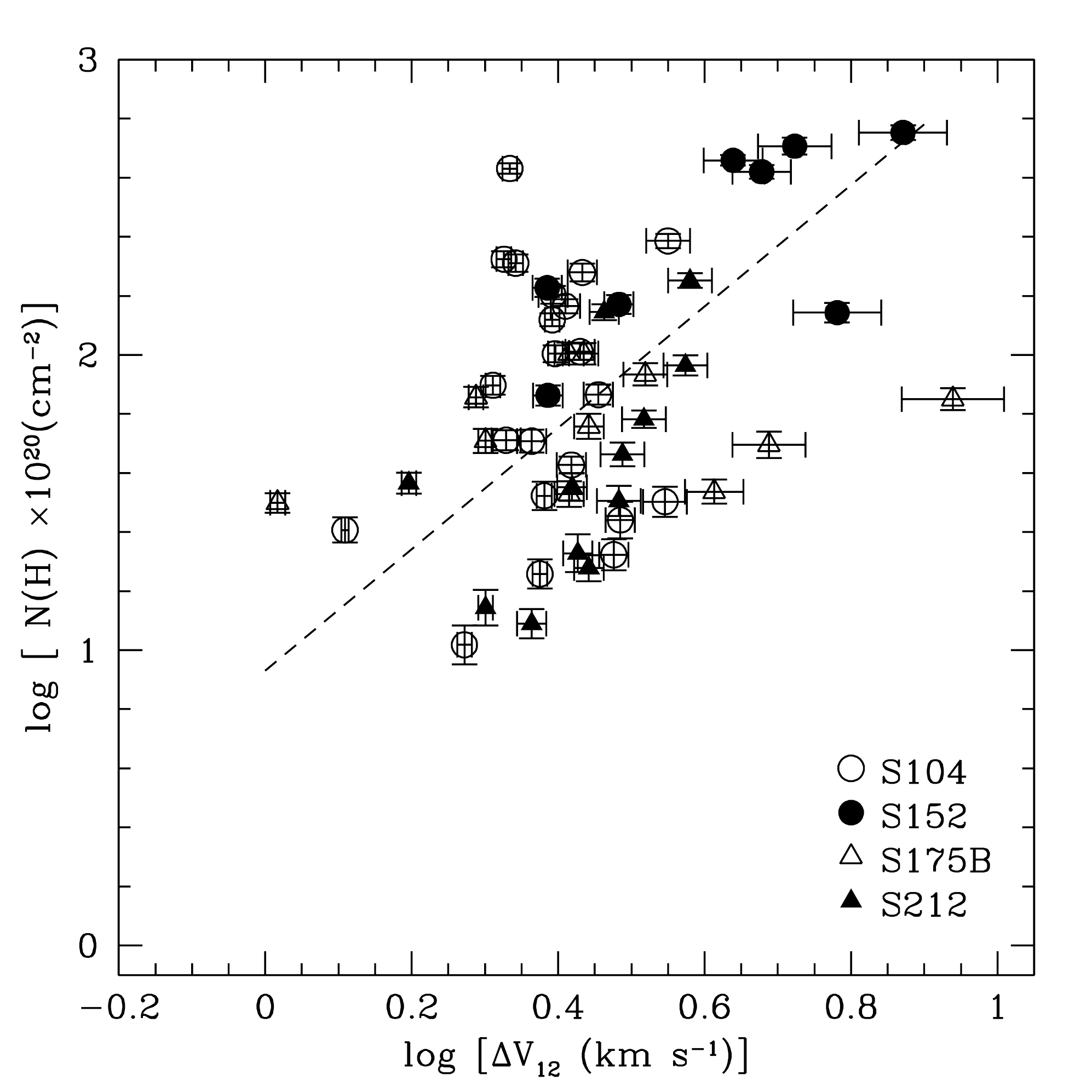}    
   \caption[LTE column density vs. $\Delta V_{12}$]{ LTE column density vs. $\Delta V_{12}$ for Type I  (left)  and Type II  (right)  sources. The dashed line shows  the bisector least-squares  fits.  The relation for Type I  sources has a slope of 2.3; the relation for Type II sources is slightly weaker (correlation coefficient of 0.42 compared to 0.53 for Type I sources) with a slope of 2.06.
   }
   \label{ncoldelv12}
\end{figure}

\begin{figure}
   \centering
   \includegraphics[width=7.5cm]{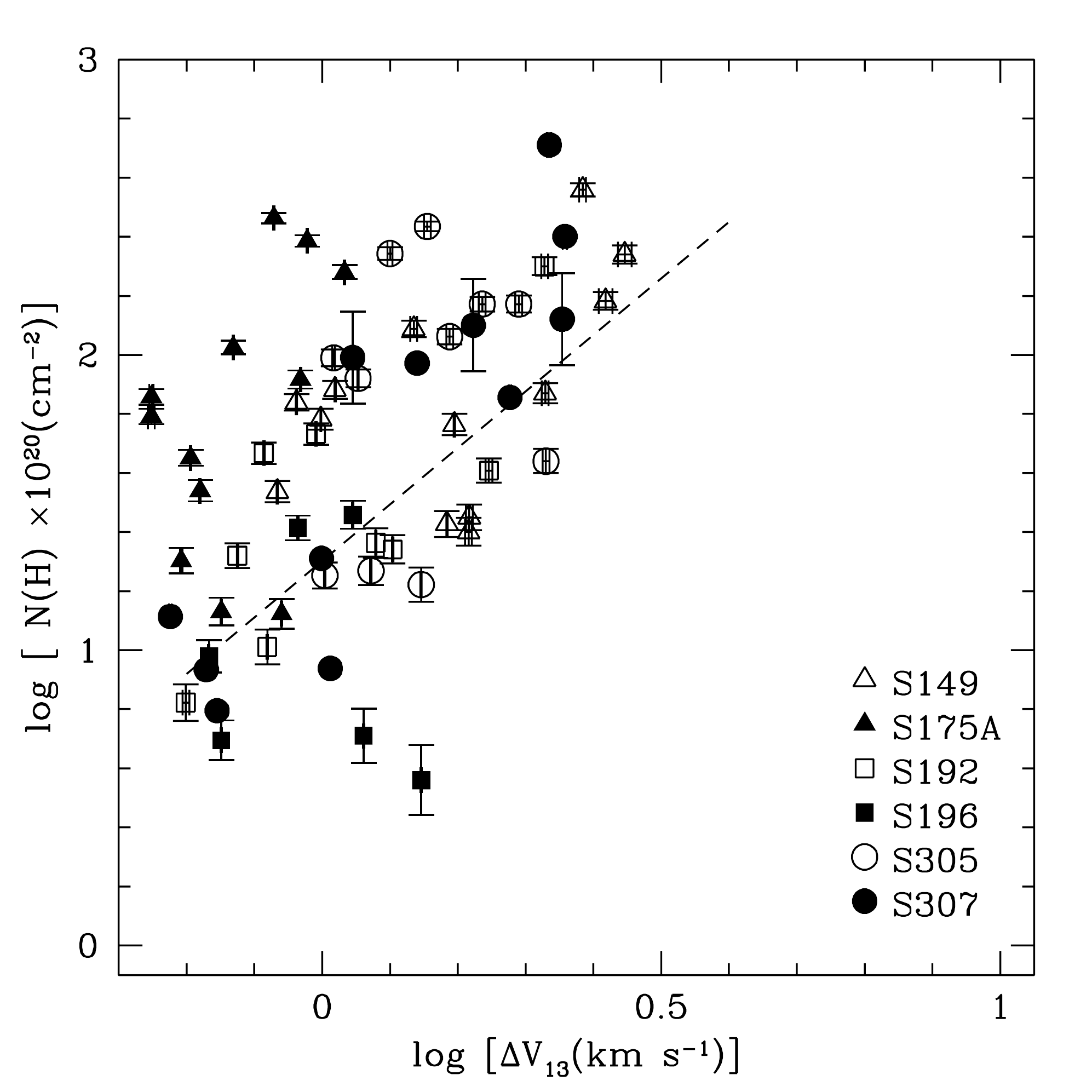}
   \includegraphics[width=7.5cm]{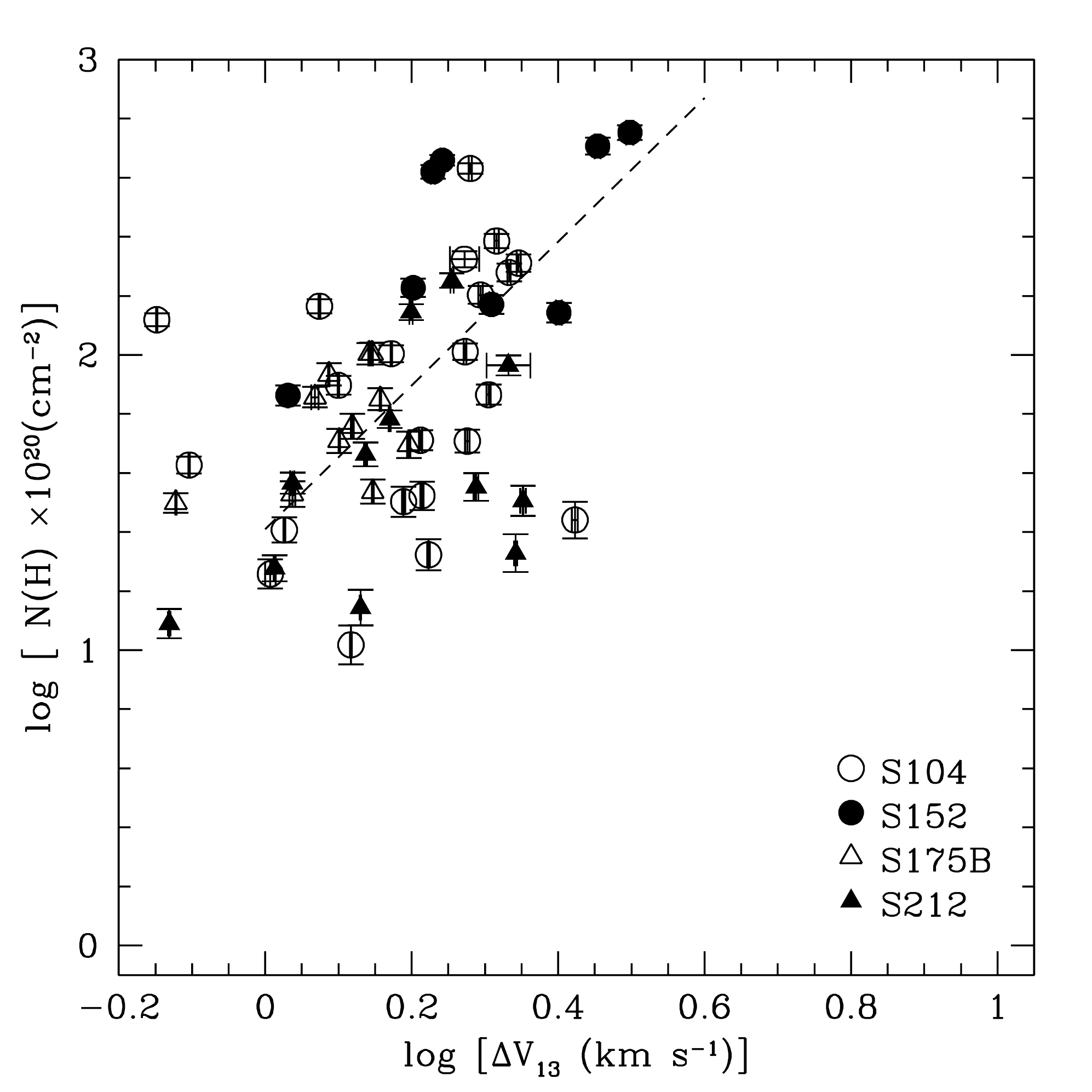}    
   \caption[LTE column density vs. $\Delta V_{13}$]{Same plot as  Figure \ref{ncoldelv12} for \tco line widths. The relation for Type I  sources has the same slope as for \co, 2.3; the relation for Type II sources is slightly weaker (correlation coefficient of 0.43 compared to 0.53 for Type I sources) with a slope of 2.43.
   }
   \label{ncoldelv13}
\end{figure}

\begin{figure}
   \centering
   \includegraphics[width=7.5cm]{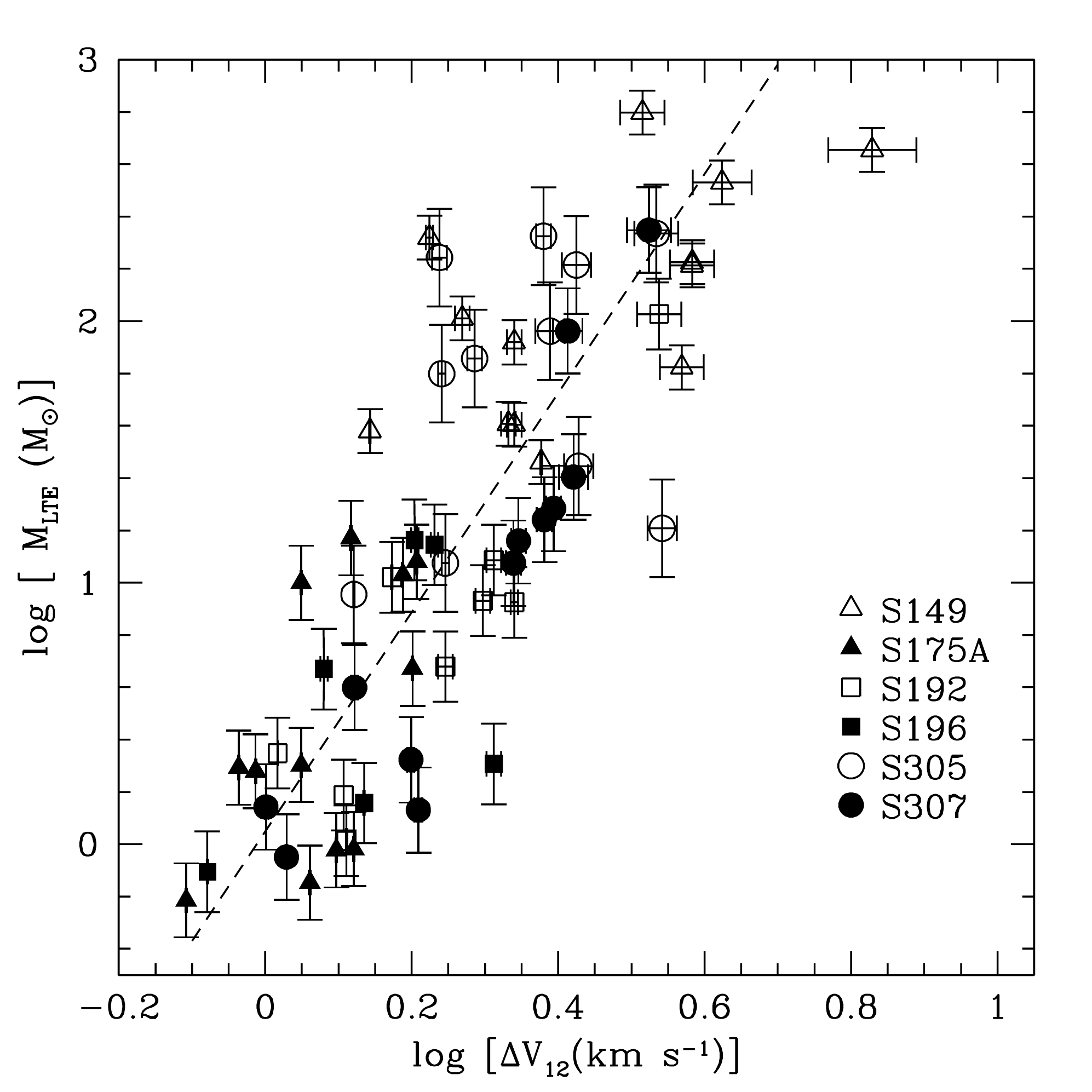}
   \includegraphics[width=7.5cm]{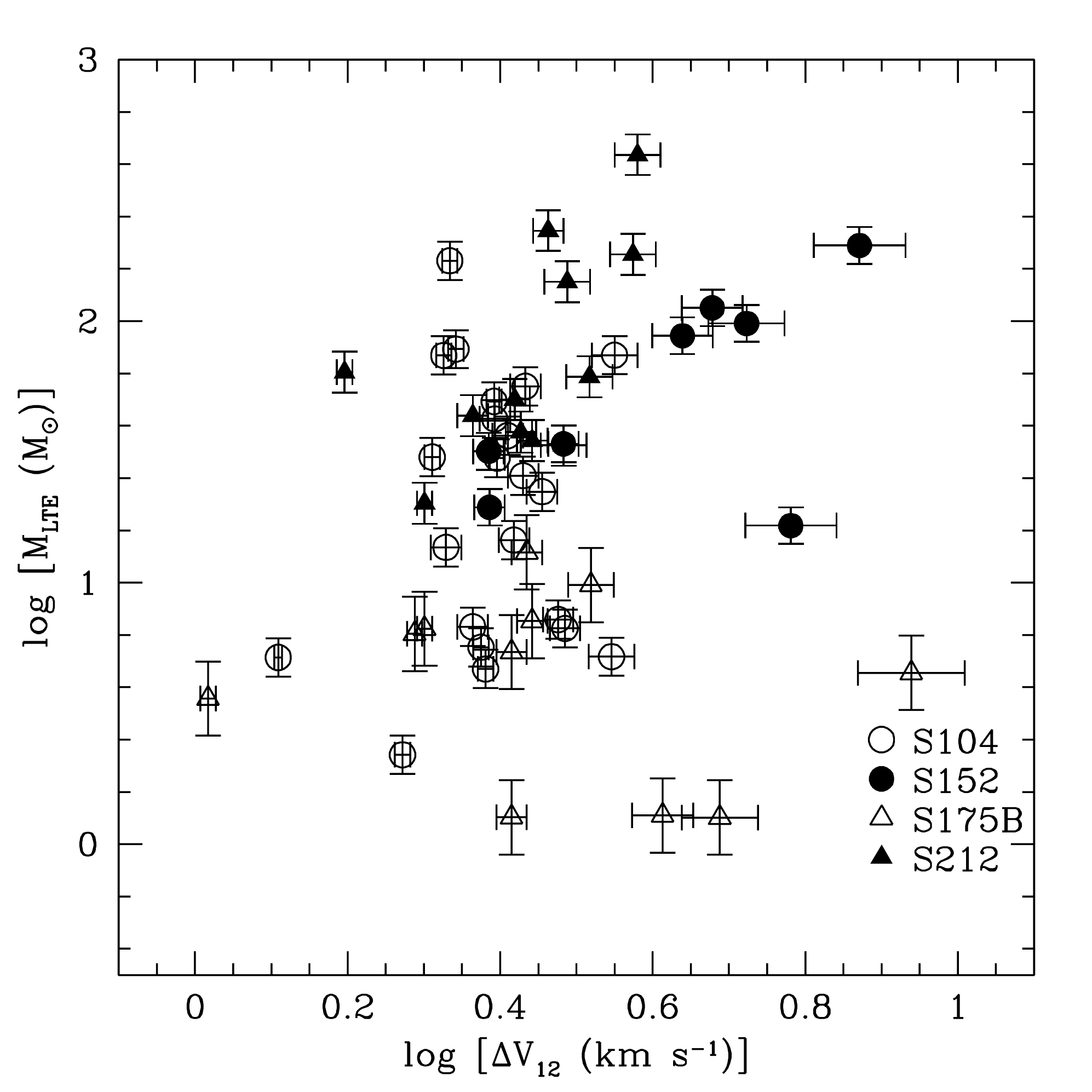}    
   \caption[$M_{LTE}$ vs. $\Delta V_{12}$ for Type I   and Type II   sources]{$M_{LTE}$ vs. $\Delta V_{12}$ for Type I  (left)  and Type II  (right)  sources. The dashed line shows the bisector least-squares  fit to Type I  clumps with a slope of 4.2. No relation is found  for Type II  sources.}
   \label{mltedelv12}
\end{figure}

\begin{figure}
   \centering
   \includegraphics[width=7.5cm]{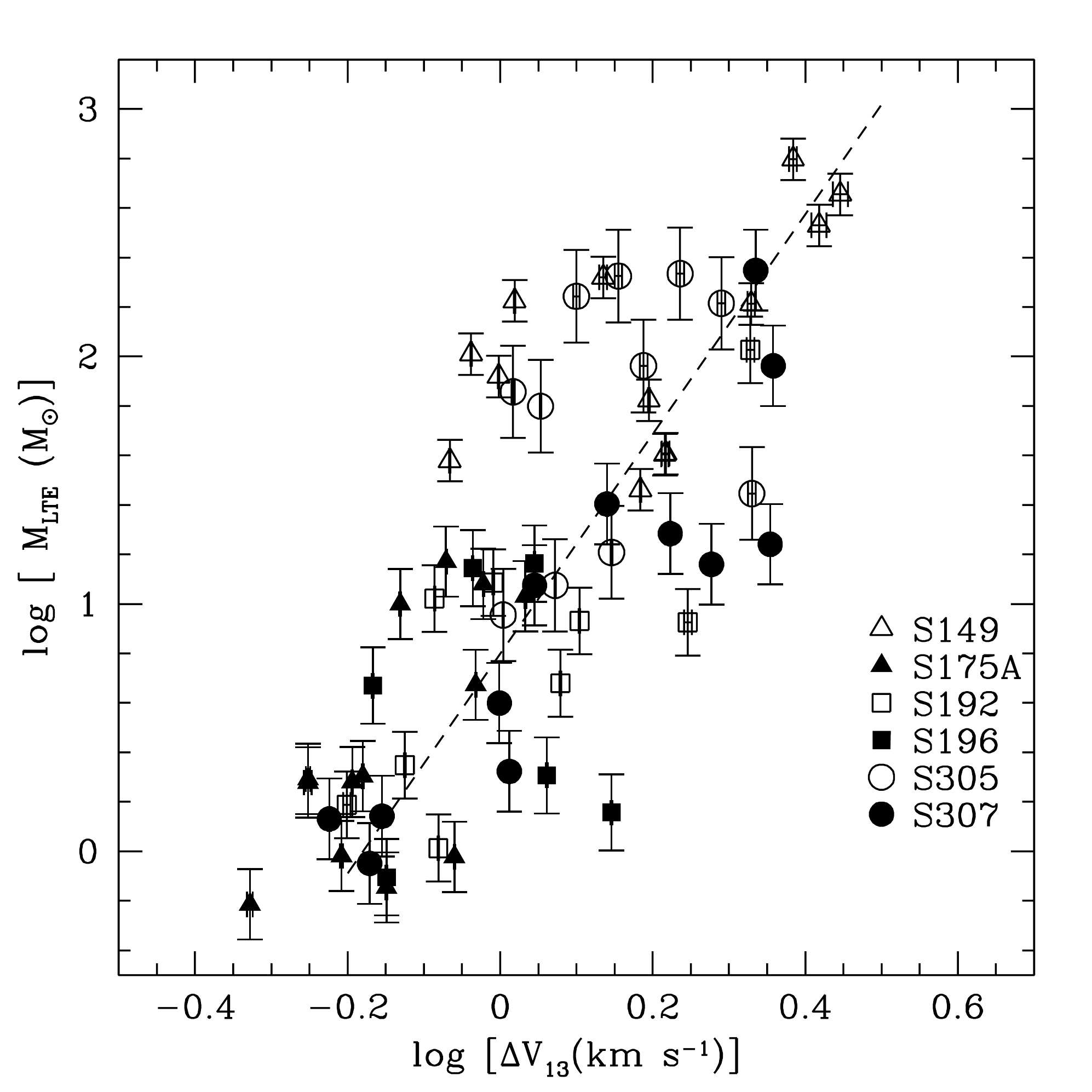}
   \includegraphics[width=7.5cm]{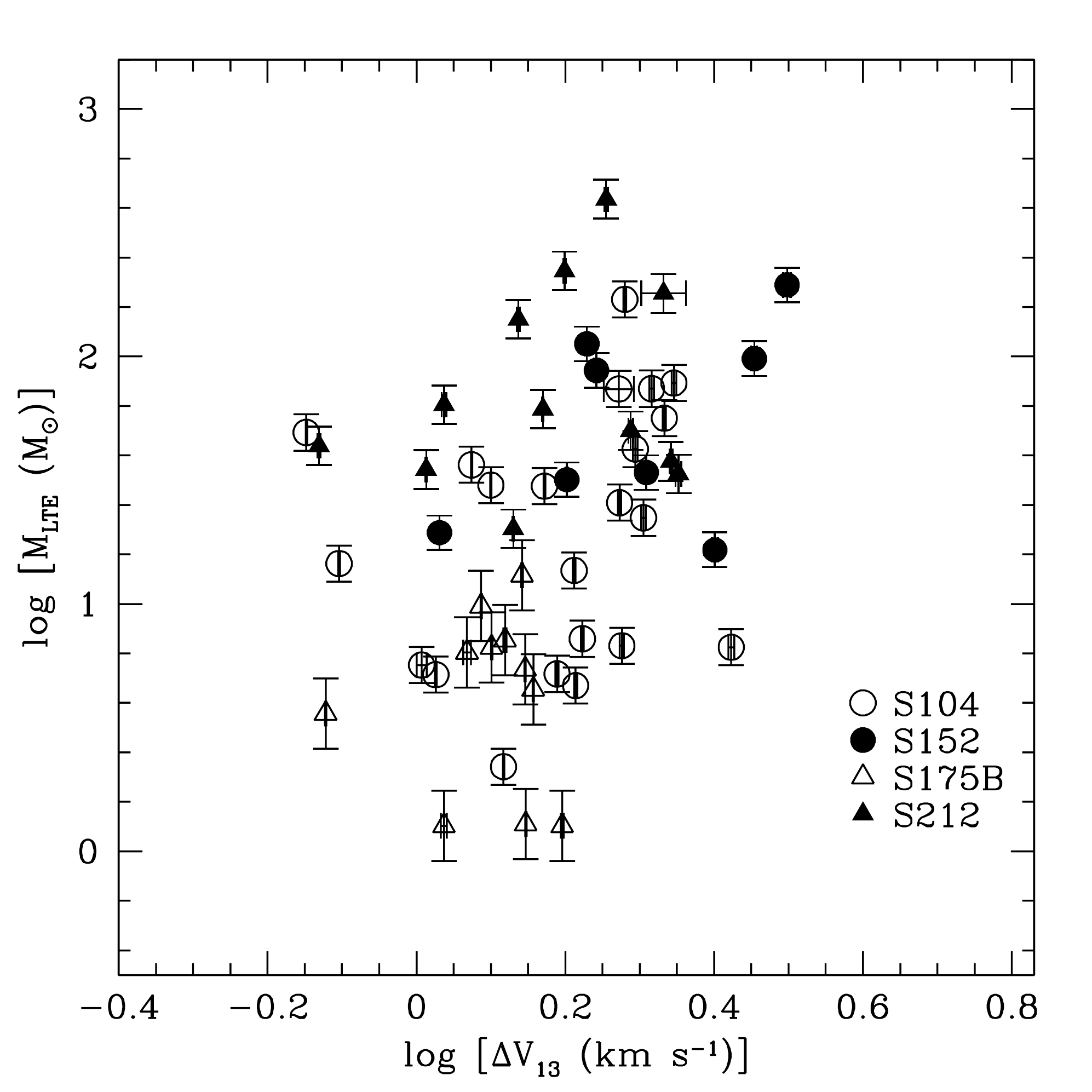}    
   \caption[$M_{LTE}$ vs. $\Delta V_{13}$ for Type I  and Type II  sources]{Similar to Figure \ref{mltedelv12} for \tco line width. The dashed line is the least-squares fits with the slope of 4.2, the same as the previous plot.  Similar to \co, no relation is found between $M_{LTE}$ and $\Delta V_{13}$ for Type II  sources.}
   \label{mltedelv13}
\end{figure}

\begin{figure}
   \centering
   \includegraphics[width=6in]{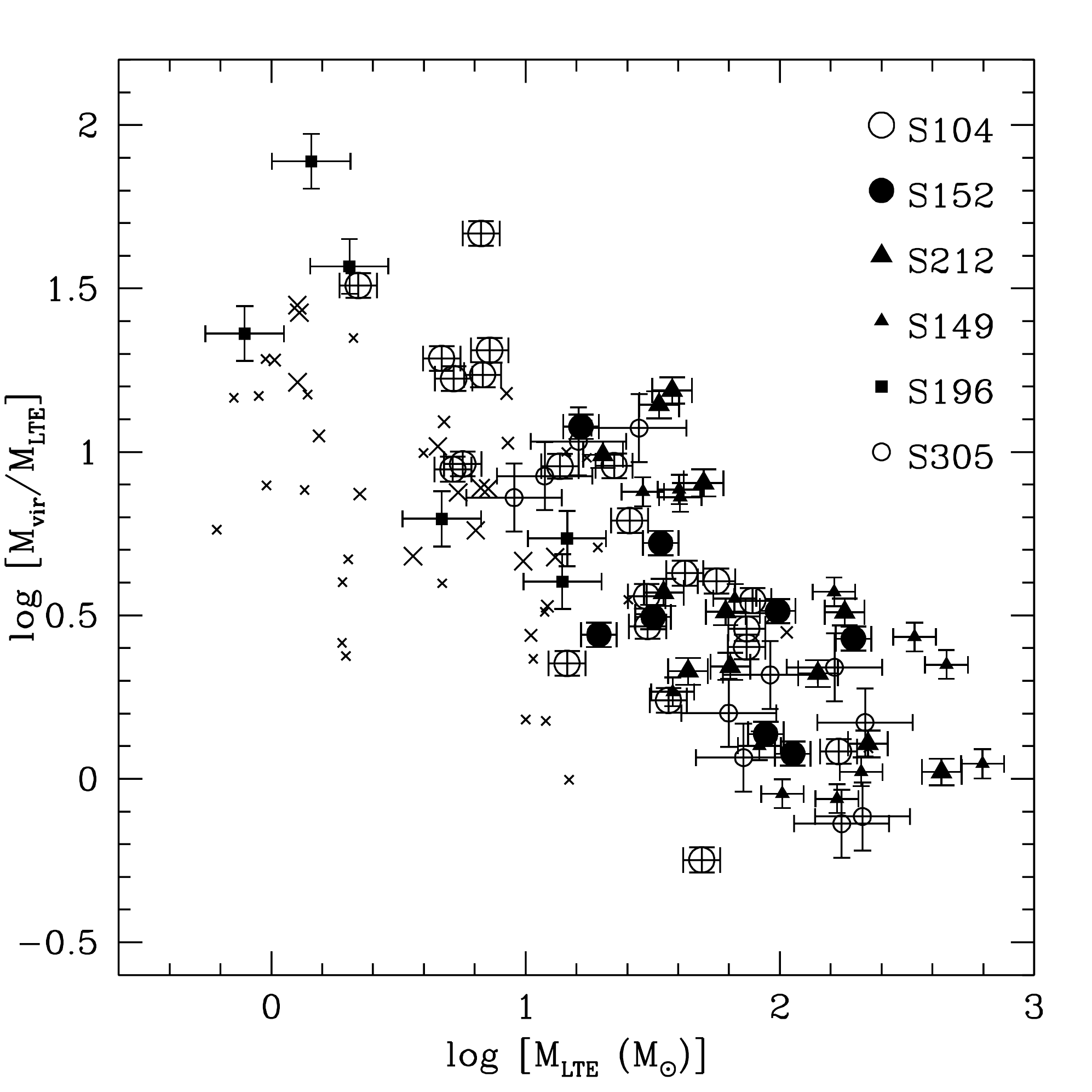} 
   \caption[$M_{vir}$/$M_{LTE}$  vs.  $M_{LTE}$]{$M_{vir}$/$M_{LTE}$  vs.  $M_{LTE}$. Filled circles present Type I  sources and open circles present Type II  sources.  To decrease the effect of varying  distances and consequently the resolution effect we have considered only sources at distances between 3.3 and 7.1 k~pc. Other sources are shown by crosses. }
   \label{mvirltelog}
\end{figure}

\begin{figure}
   \centering
   \includegraphics[width=6in]{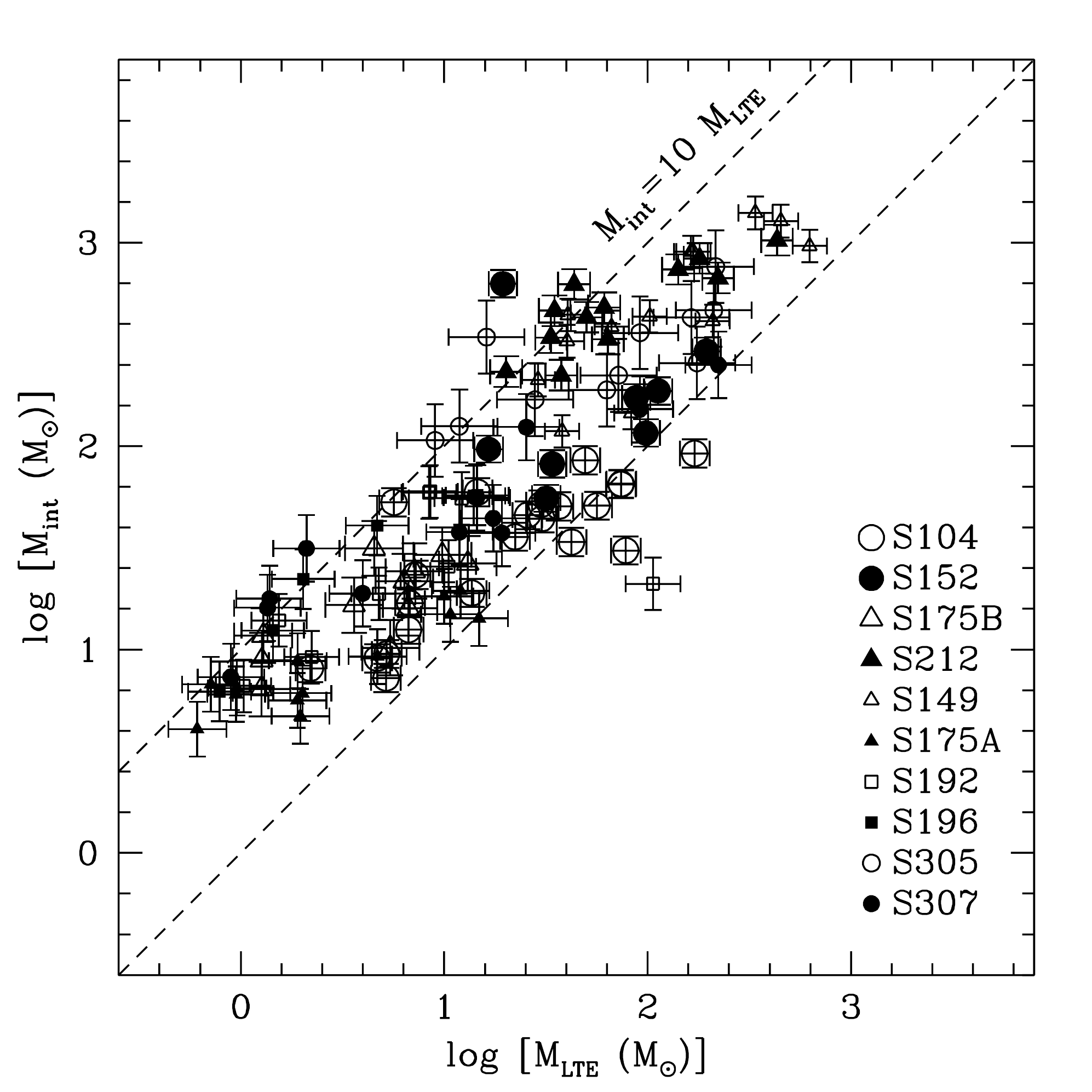} 
   \caption[Velocity integrated mass plotted vs. LTE mass]{Velocity integrated mass plotted vs. LTE mass. Filled circles represent Type I  sources and open circles represent Type II  sources. The dashed lines are $M_{int}=M_{LTE}$ and $M_{int }= 10 M_{LTE}$. }
   \label{mintlte}
\end{figure}

\begin{figure}
   \centering
   \includegraphics[width=6in]{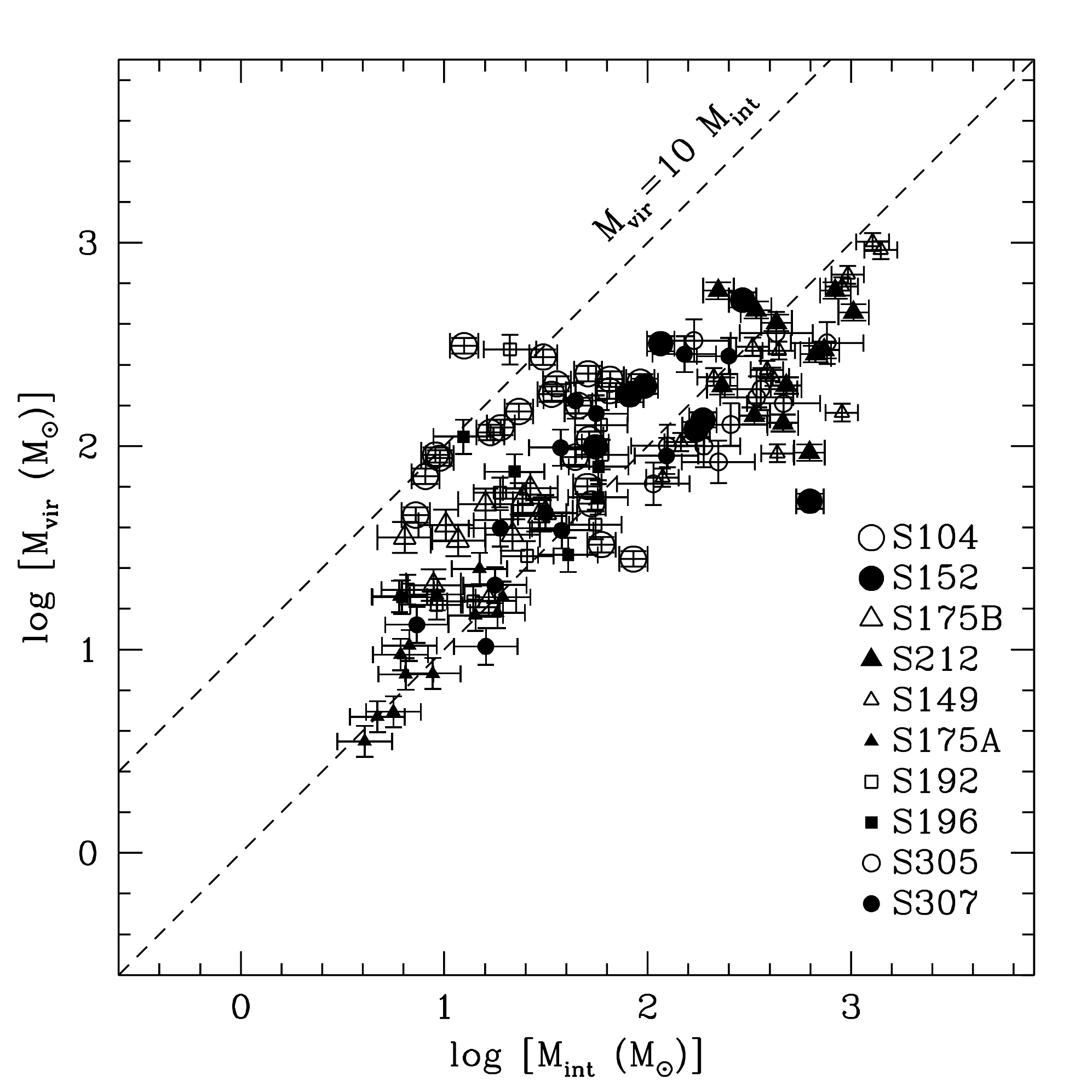} 
   \caption[Virial mass plotted vs. velocity integrated mass]{ virial mass plotted vs. velocity integrated mass. Filled circles represent Type I  sources and open circles represent Type II  sources.  The dashed lines are $M_{vir}=M_{int}$ and $M_{vir }= 10 M_{int}$.  Virial mass tends to equal to $ M_{int}$ for massive clumps and they are equal at  M $\sim 100 M_\odot$.  }
   \label{mvirint}
\end{figure}

\begin{figure}
   \centering
   \includegraphics[width=3in]{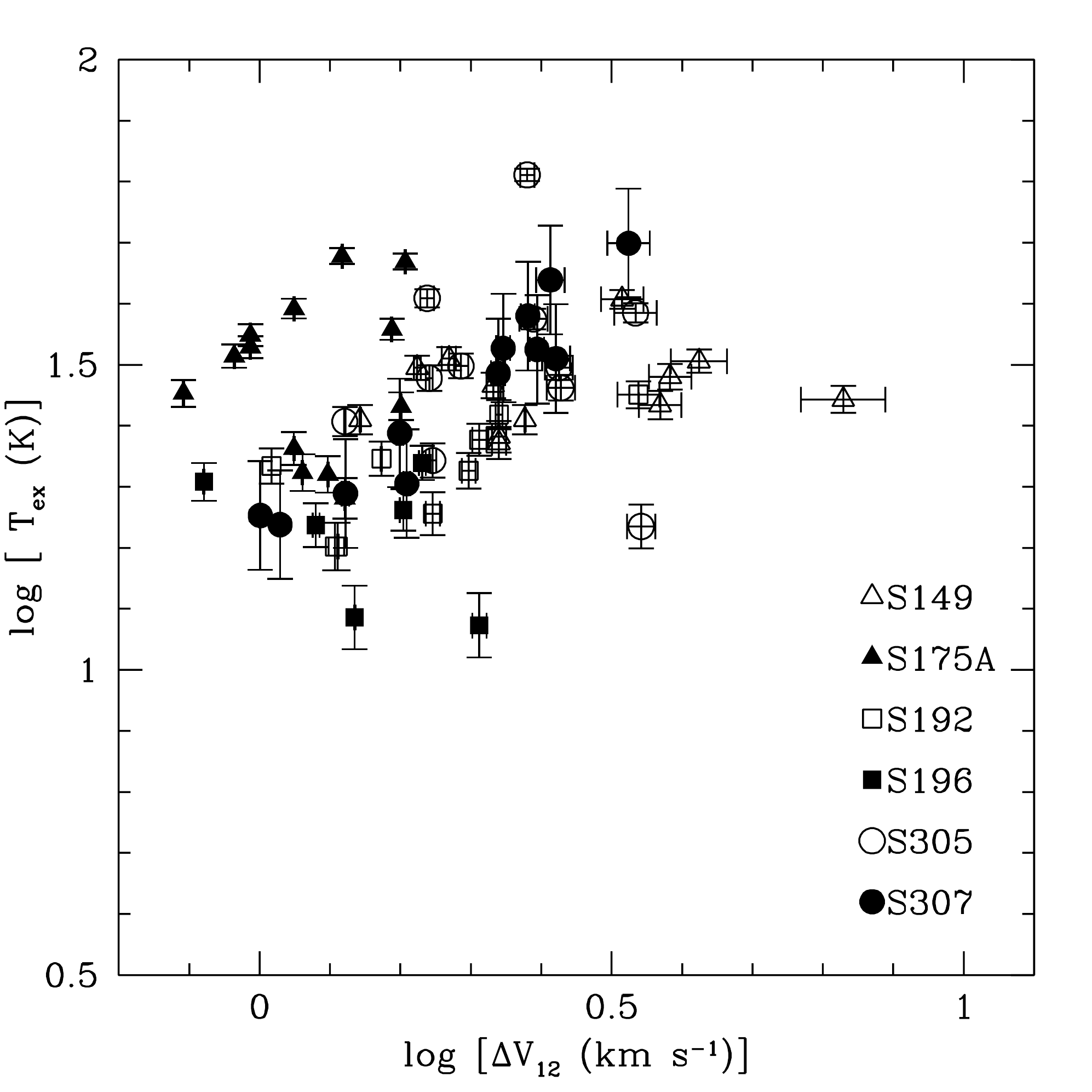} 
   \includegraphics[width=3in]{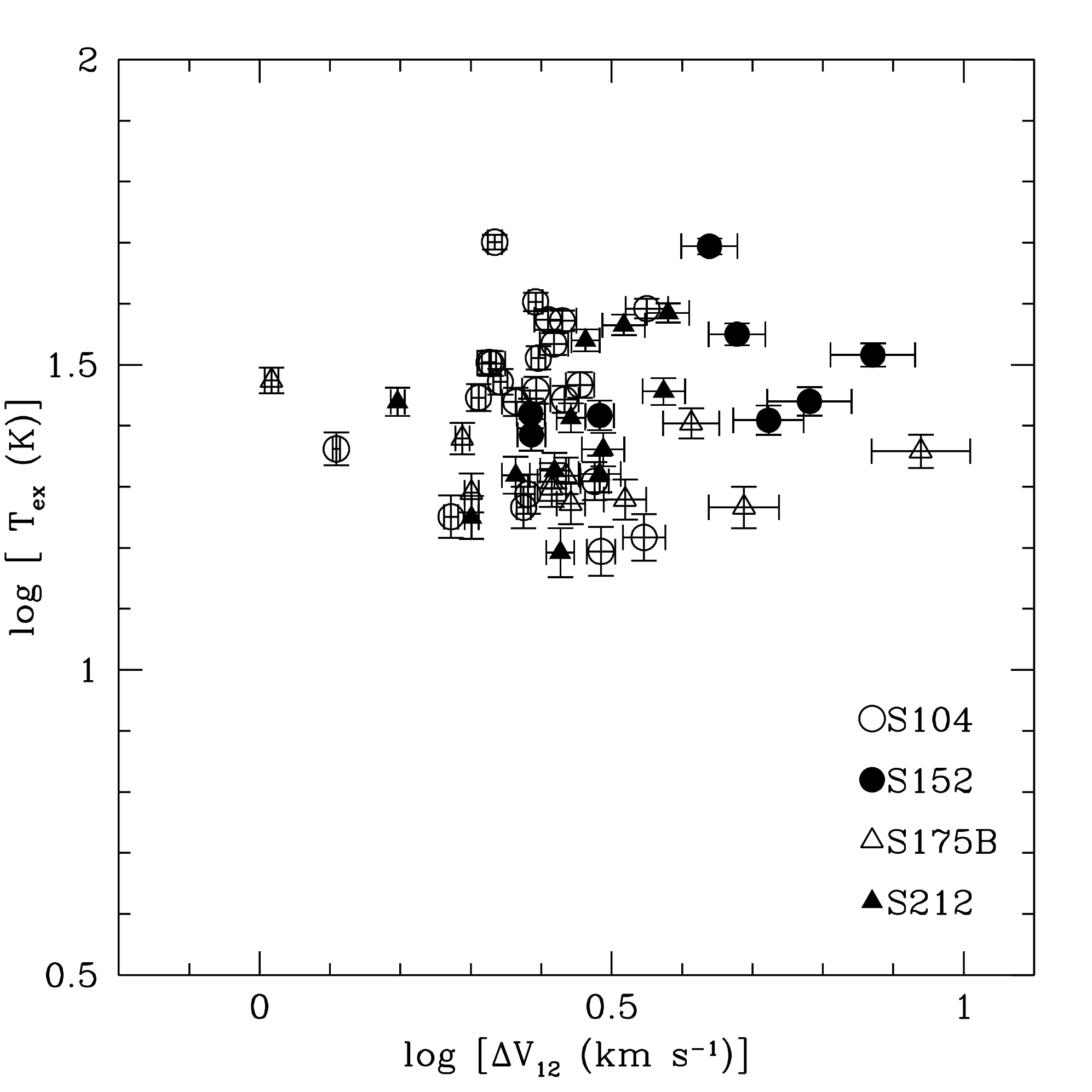} 
   \caption[Excitation temperature vs.  $\Delta V_{12}$]{ Excitation temperature vs.  $\Delta V_{12}$ for Type I  (left panel) and Type II  (right panel) sources.  No relation is found for Type I or Type II sources.}
   \label{texdelv12}
\end{figure}

\begin{figure}
   \centering
   \includegraphics[width=3in]{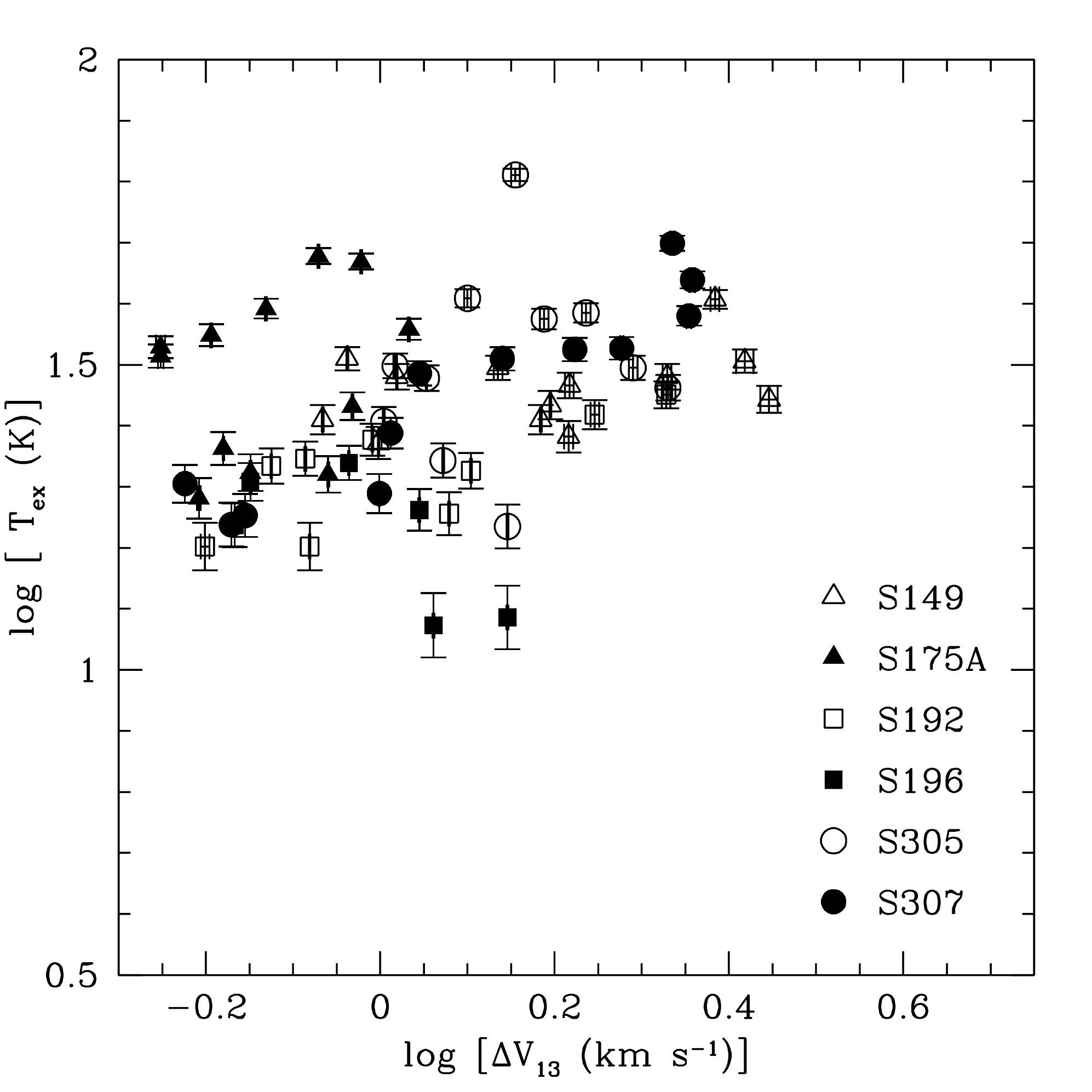} 
   \includegraphics[width=3in]{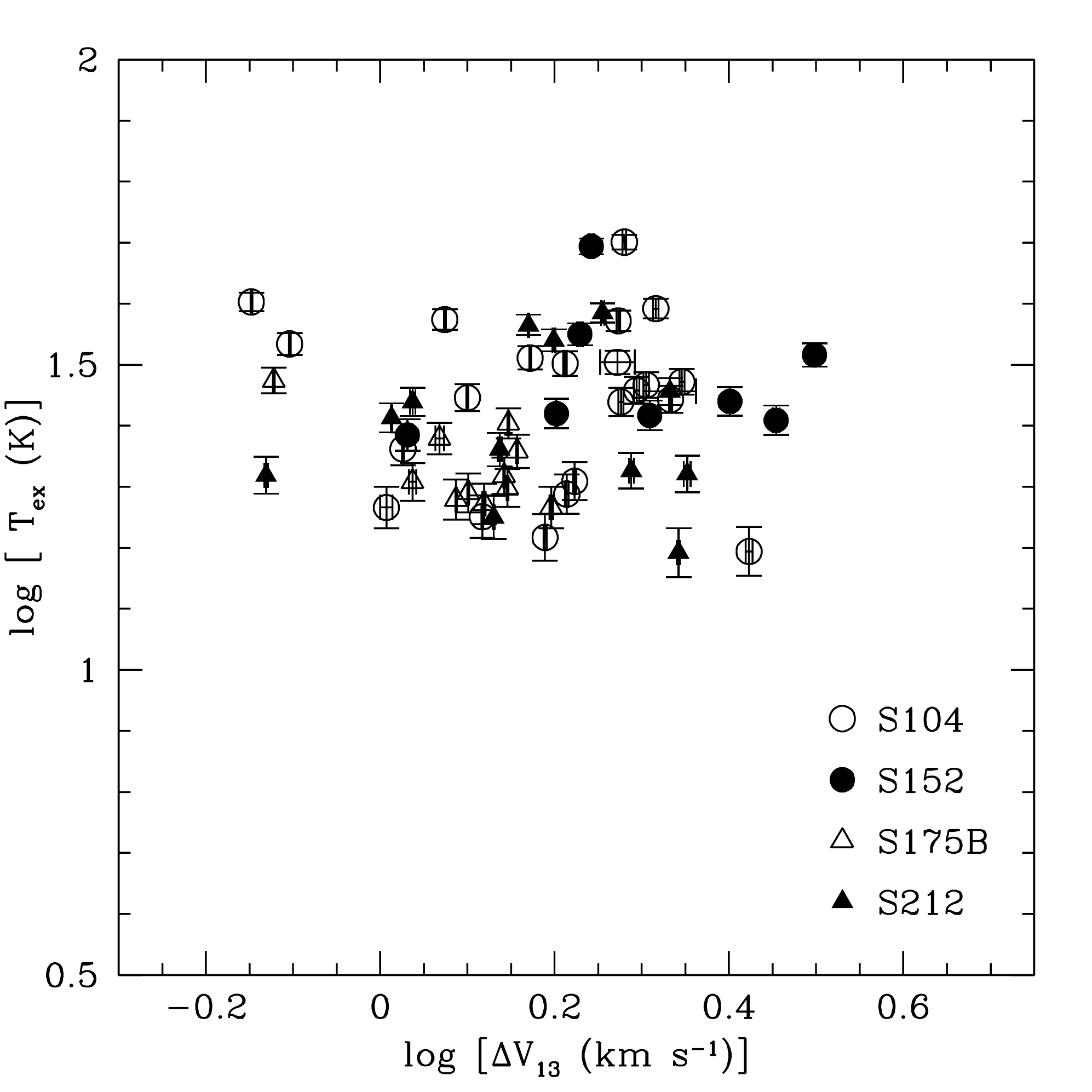} 
   \caption[Excitation temperature vs.  $\Delta V_{13}$]{ Same as Figure \ref{texdelv12} for $\Delta V_{13}$. Similarly no relation is found.}
   \label{texdelv13}
\end{figure}

\begin{figure}
   \centering
   \includegraphics[width=3in]{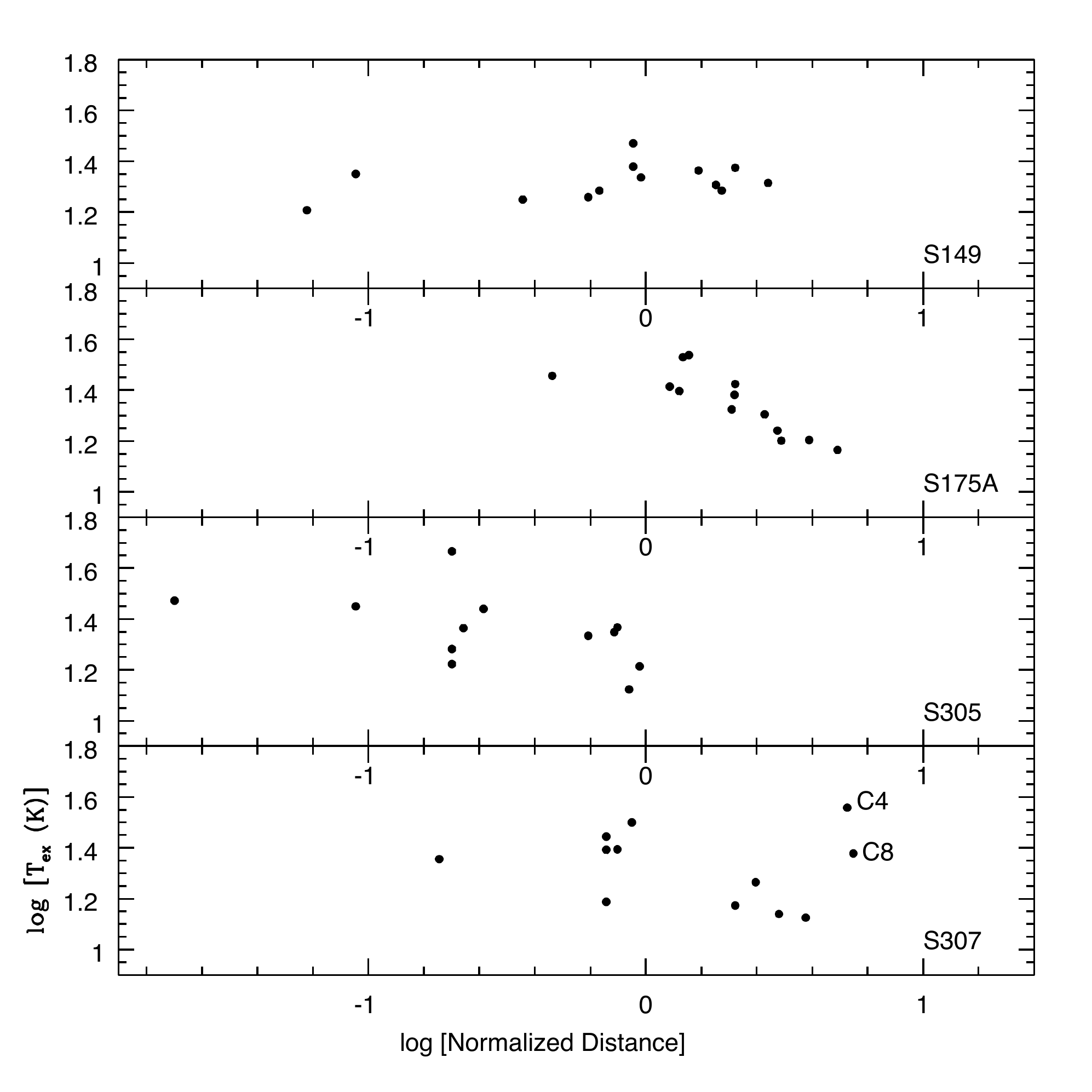} 
   \includegraphics[width=3in]{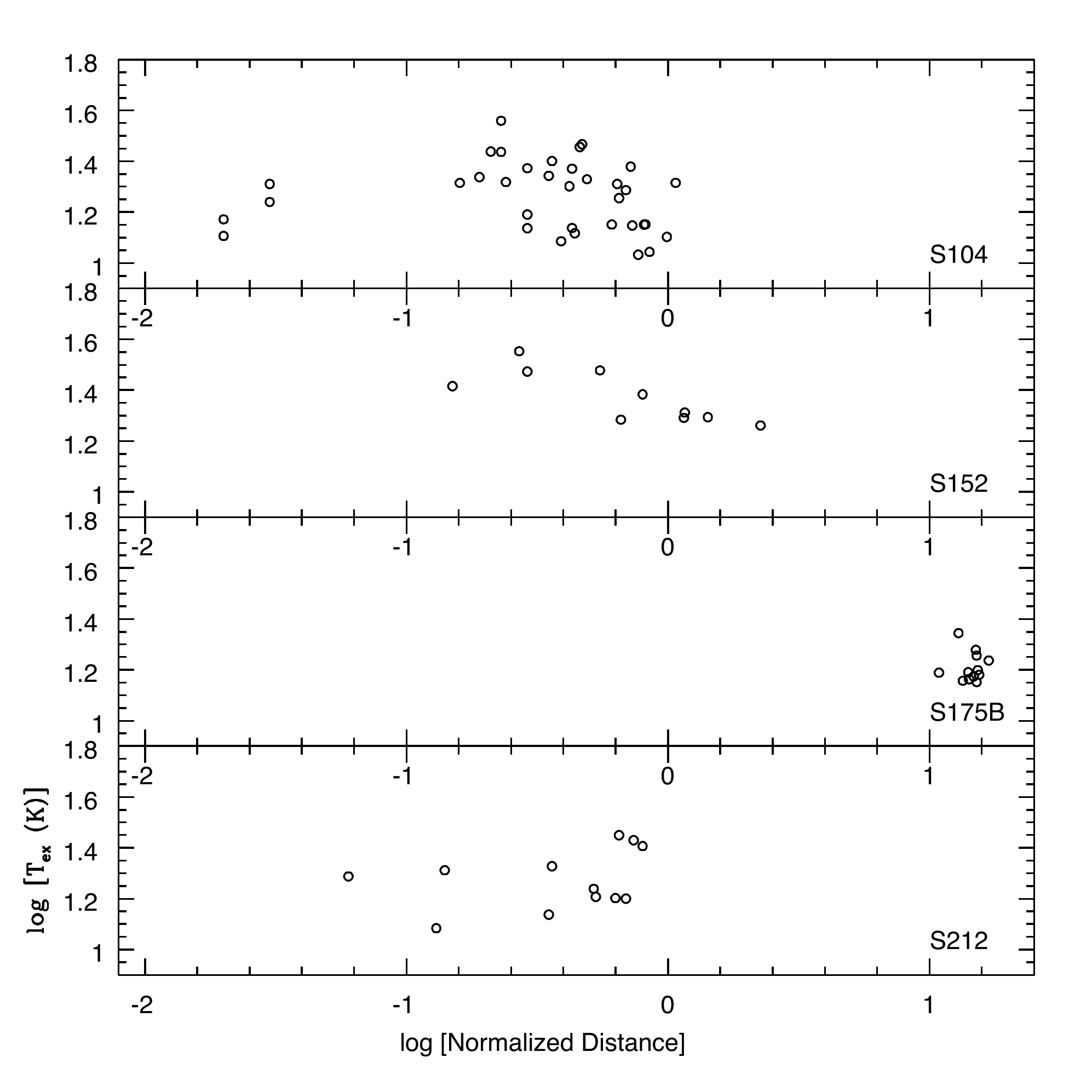} 
   \caption[Excitation temperature vs. normalized distance from H~II region]{Excitation temperature vs. normalized distance from H~II region for Type I  (left panel) and $^{13}$CO(right panel) sources.}
   \label{texndist}
\end{figure}

\begin{figure}
   \centering
   \includegraphics[width=3in]{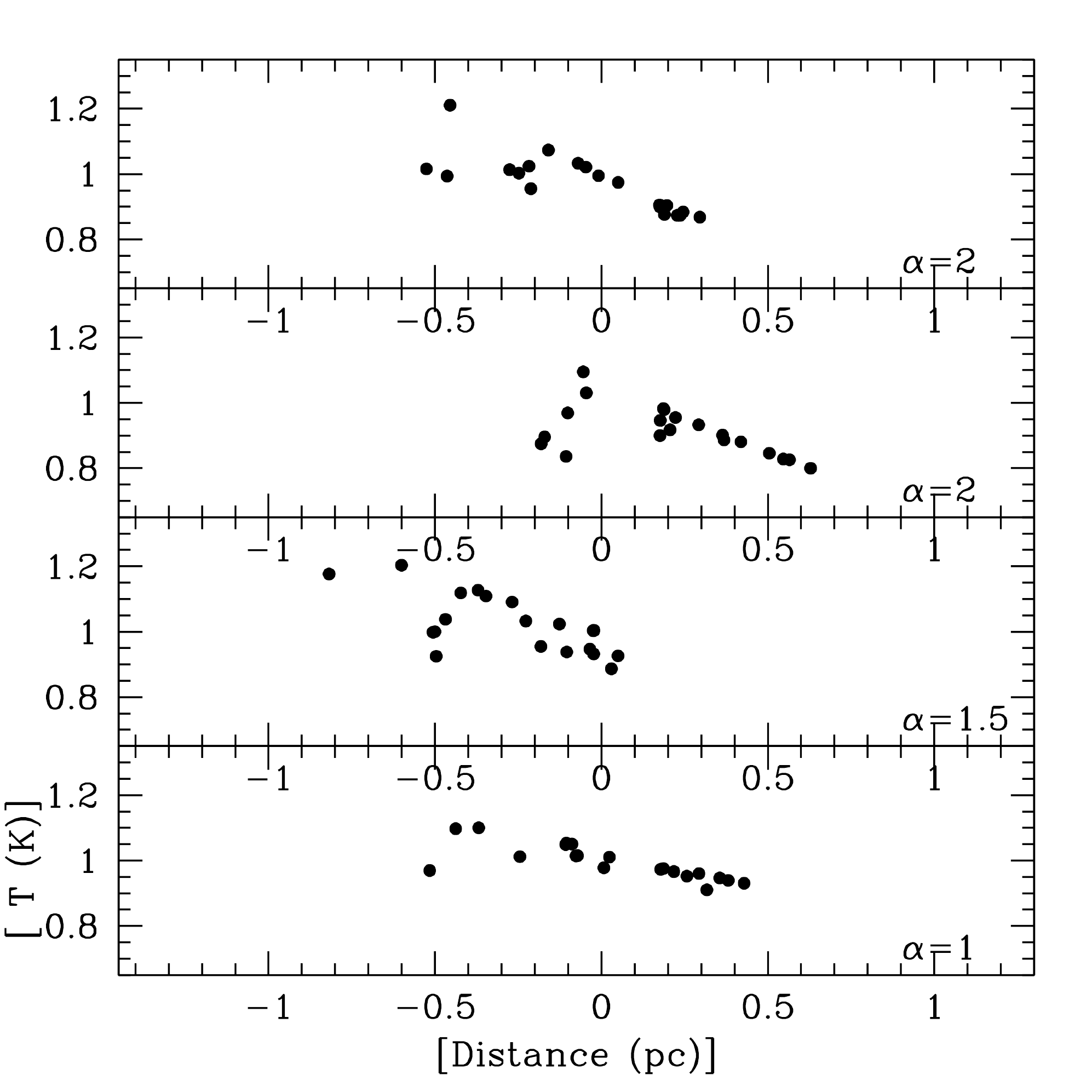} 
      \caption[Simulated temperature variation from a heating source]{Simulated temperature variation from a heating source with  similar  physical  conditions as our cloud samples with different  luminosity decrease power law index, $\alpha$.} 
   \label{tsim}
\end{figure}

\begin{figure}
   \centering
   \includegraphics[width=7.5cm]{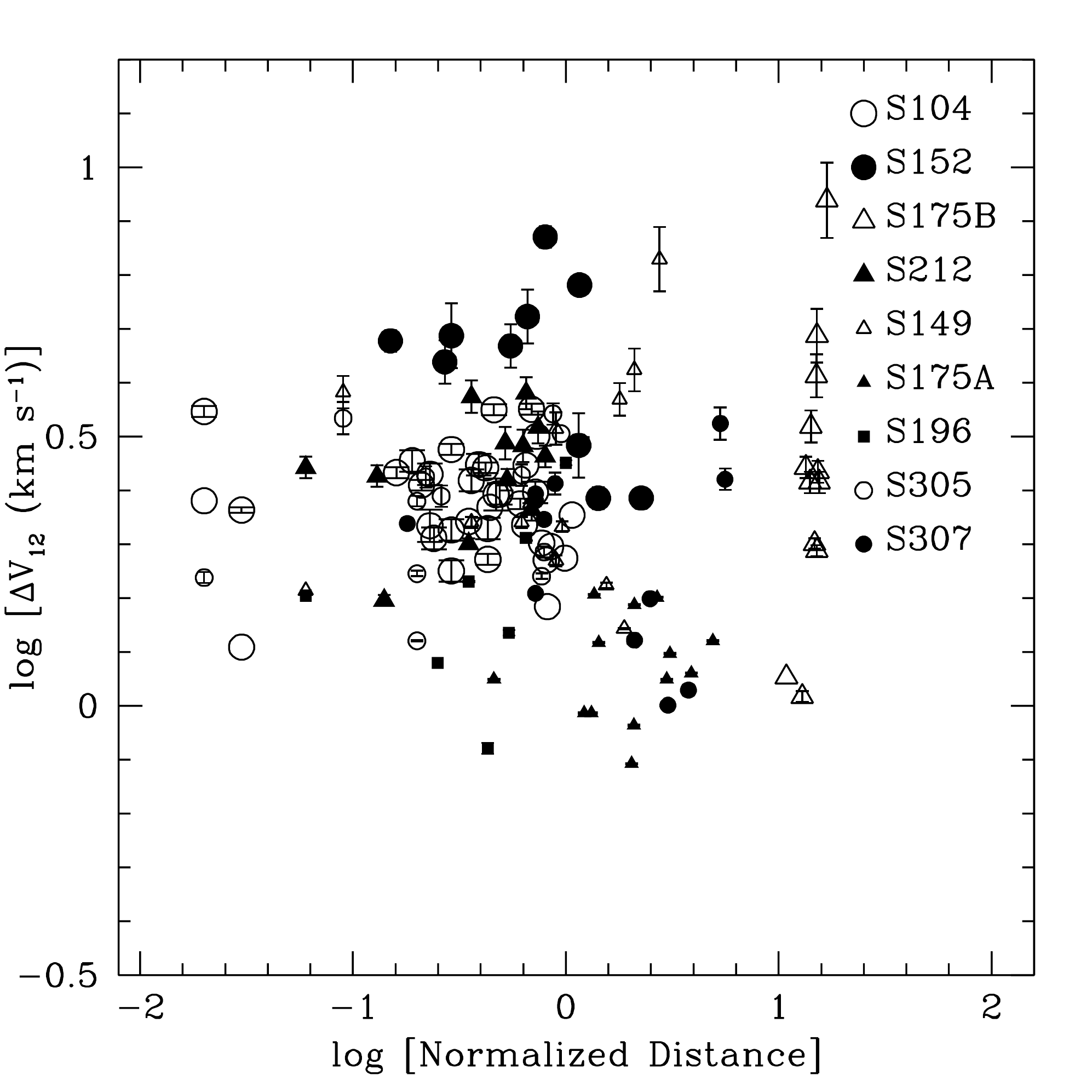} 
   \includegraphics[width=7.5cm]{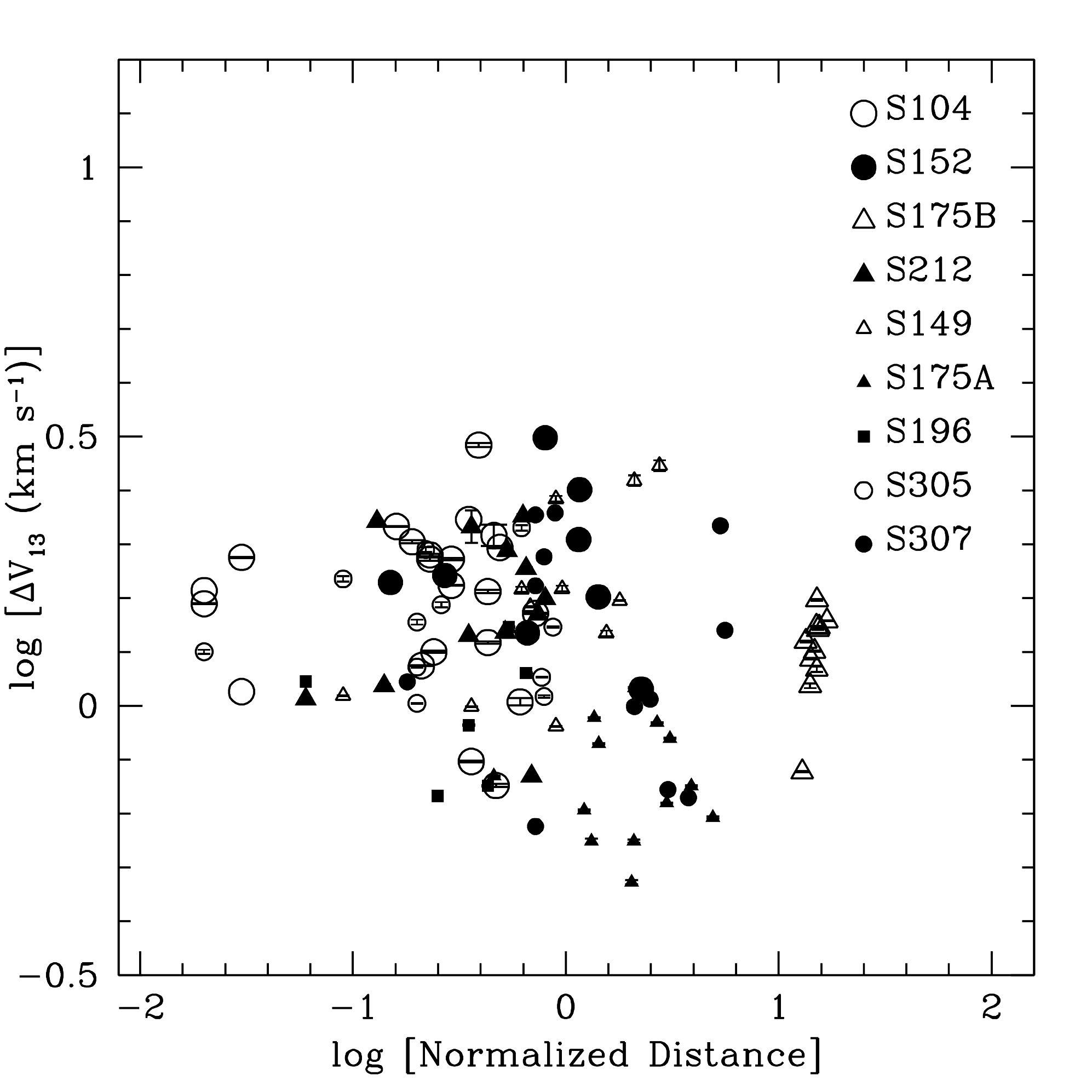}
   \caption[Line width vs. normalized distance from H~II region]{Line width vs. normalized distance from H~II region for $\Delta V_{12}$ (left)  and $\Delta V_{13}$ (right).  }
   \label{delvndist}
\end{figure}

\clearpage

\begin{deluxetable}{llllccccc}
\tabletypesize{\scriptsize}
\tablecaption{Properties of selected regions \label{sample}}
\tablewidth{0pt}
\tablehead{
\colhead{Source}& \colhead{RA} & \colhead{Dec} &
\colhead{Distance} & \colhead{$V_{LSR}$}& Diameter& Diameter& \colhead{Exciting Star} 
\\
\colhead{}&\colhead{J(2000)}&\colhead{J(2000)}&\colhead{(kpc)}&\colhead{(kms$^{-1}$)}&
\colhead{(arcmin)}&\colhead{(pc)}&
}
\startdata

S104&20:17:42&36:45:30&3.3$\pm$0.3&0.0&7&6.7 & O6V  \tablenotemark{a} \\
S148-S149&22:56:22&58:31:29&5.6$\pm$0.6& -53.1&1&1.6&B0V \tablenotemark{a}\\
S152&22:58:41&58:47:06&2.39$\pm$0.21&-50.4&2&1.4&O9V \tablenotemark{b}\\
S175A&0:27:04&64:43:35&1.09$\pm$0.21&-49.6&2&0.63&B1.5V \tablenotemark{b}\\
S175B&0:26:25&64:52:36&1.09$\pm$0.21&-49.6&2&0.63&B1.5V \tablenotemark{b}\\
S192-S194&2:47:30&61:56:33&2.96$\pm$0.54&-46.3&1&0.86&B2.5V \tablenotemark{b}\\
S196&2:51:41&62:12:19&4.7$\pm$1.0&-45.1&4&5.4&\nodata \tablenotemark{d}\\
S212&4:40:56&50:27:47&7.1$\pm$0.7&-35.3&5&10.3&O6 \tablenotemark{a}\\
S288 &7:8:39&-4:18:41&3.0$\pm$1.2&56.7&1&0.87&B1 \tablenotemark{c}\\
S305&7:30:13&-18:31:50&5.2$\pm$1.4&44.1&4&6.1&O9.5 \tablenotemark{c}\\
S307&7:35:33&-18:45:55&2.2$\pm$0.5&46.3&6&3.8&O9 \tablenotemark{c}\\
\enddata
\tablenotetext{a}{Data about exciting star and distance from  Caplan et~al. 2000 }
\tablenotetext{b}{Data about exciting star and distance from Russeil et~al. 2007 }
\tablenotetext{c}{Data about exciting star and distance from Moffat et al. 1997 }
\tablenotetext{d}{No exciting star identified  for this H~II region. The kinematic distance is reported here. Note that S196 is close to S192/S193 in position and velocity and is likely to  be at the same distance.}
\end{deluxetable}

\begin{deluxetable}{lccc}
\tabletypesize{\scriptsize}
\tablecaption{Comparison with previous works \label{comparison}}
\tablewidth{0pt}
\tablehead{
\colhead{Data}& \colhead{Slope} & \colhead{Sigma} &
\colhead{Correlation Coefficient} 
}
\startdata
Kim \& Koo (2003)& 0.35$\pm$0.06& 0.10&0.79\\
Saito et al. (2006)& 0.44$\pm$0.12& 0.16&0.57\\
Yamamoto et al. (2006)& 0.29$\pm$ 0.07& 0.15&0.44\\
Type I Sources $^{12}$CO & 0.51$\pm$0.06& 0.14&0.70\\
Type I Sources $^{13}$CO & 0.47$\pm$0.07& 0.15&0.67\\
Type II Sources & -0.09$\pm$0.09& 0.17&-0.12\\
\enddata
\end{deluxetable}

\clearpage
\begin{deluxetable}{cccccccc}
\tabletypesize{\scriptsize}
\tablecaption{Physical parameters  measured for the entire sample \label{tbl2}}
\tablewidth{0pt}
\tablehead{
\colhead{ID}  & \colhead{RA} & \colhead{DEC}&
\colhead{$R_{e}$} & \colhead{$^{12}T_b$} & \colhead{$^{13}T_b$}&\colhead{V$_{12}$}&\colhead{V$_{13}$}\\
 \colhead{}&\colhead{}&\colhead{}&\colhead{(pc)}&\colhead{(K)}&\colhead{(K)}&\colhead{(km s$^{-1}$)}&\colhead{(km s$^{-1}$)}
}
\startdata
{\bf S104}&&&&&&\\
\hspace{1cm}... C1 & 20:17:26.0 & +36:42:34.0 & 0.40$\pm$0.04 & 13.3 & 3.2 & 0.81 & 0.58\\
\hspace{1cm}... C2 & 20:17:34.4 & +36:46:39.5 & 0.35$\pm$0.03 & 10.5 & 2.1 & -2.85 & -2.32\\
\hspace{1cm}... C3 & 20:18:01.8 & +36:45:10.1 & 0.38$\pm$0.03 & 20.5 & \nodata & -0.1 & \nodata \\
\hspace{1cm}... C4 & 20:17:56.6 & +36:45:23.8 & 0.45$\pm$0.04 & 44.7 & 18.6 & 0.61 & -0.42  \\
\hspace{1cm}... C5 & 20:17:23.9 & +36:45:56.8 & 0.28$\pm$0.03 & 12.6 & \nodata & 1.83 & \nodata  \\
\hspace{1cm}... C6 & 20:17:50.7 & +36:46:54.5 & 0.29$\pm$0.03 & 11.4 & 3.7 & -1.42 & -1.33  \\
\hspace{1cm}... C7 & 20:17:53.6 & +36:48:18.7 & 0.42$\pm$0.04 & 28.8 & 6.7 & 1.42 & -0.66  \\
\hspace{1cm}... C8 & 20:17:57.8 & +36:44:34.9 & 0.44$\pm$0.04 & 34.7 & 17.1 & 0 & 0.66  \\
\hspace{1cm}... C9 & 20:17:30.3 & +36:47:49.3 & 0.39$\pm$0.04 & 22.3 & 10.7 & 0.91 & 0.66  \\
\hspace{1cm}... C10 & 20:17:57.2 & +36:43:52.8 & 0.40$\pm$0.04 & 33.6 & 12.3 & 0.81 & 0.87  \\
\hspace{1cm}... C11 & 20:17:57.7 & +36:48:11.9 & 0.40$\pm$0.04 & 27.0 & 8.4 & 0.4 & 0.54  \\
\hspace{1cm}... C12 & 20:17:29.9 & +36:42:55.2 & 0.42$\pm$0.04 & 26.5 & 12.5 & 1.62 & 1.25  \\
\hspace{1cm}... C13 & 20:17:35.1 & +36:42:34.6 & 0.36$\pm$0.03 & 32.0 & 12.9 & 1.22 & 1.37  \\
\hspace{1cm}... C14 & 20:17:40.9 & +36:42:20.9 & 0.36$\pm$0.03 & 31.9 & 6.6 & 2.23 & 1.53  \\
\hspace{1cm}... C15 & 20:17:27.4 & +36:47:21.1 & 0.44$\pm$0.04 & 22.6 & 8.2 & 1.62 & 1.7  \\
\hspace{1cm}... C16 & 20:17:30.3 & +36:48:45.3 & 0.37$\pm$0.03 & 23.3 & 10.0 & 1.01 & 1.16  \\
\hspace{1cm}... C17 & 20:17:54.9 & +36:43:17.7 & 0.37$\pm$0.03 & 26.4 & 4.3 & 1.32 & 1.45  \\
\hspace{1cm}... C18 & 20:17:36.1 & +36:47:07.6 & 0.42$\pm$0.04 & 15.1 & 2.3 & 3.66 & 3.49  \\
\hspace{1cm}... C19 & 20:17:45.0 & +36:42:35.2 & 0.4$\pm$0.04 & 23.9 & 5.1 & 4.98 & 4.9  \\
\hspace{1cm}... C20 & 20:17:46.7 & +36:44:48.3 & 0.33$\pm$0.3 & 12.7 & 1.6 & 5.48 & 5.27  \\
\hspace{1cm}... C21 & 20:17:42.6 & +36:42:56.0 & 0.26$\pm$0.02 & 22.1 & 4.1 & 5.99 & 6.52  \\
\hspace{1cm}... C22 & 20:17:42.1 & +36:41:53.0 & 0.31$\pm$0.03 & 21.5 & \nodata & 2.13 & \nodata  \\
\hspace{1cm}... C23 & 20:17:51.3 & +36:44:13.5 & 0.27$\pm$0.02 & 14.2 & 3.6 & 6.09 & 5.85  \\
\hspace{1cm}... C24 & 20:17:52.4 & +36:45:02.6 & 0.32$\pm$0.03 & 17.7 & 3.9 & 5.89 & 6.02  \\
\hspace{1cm}... C25 & 20:17:50.1 & +36:48:32.5 & 0.44$\pm$0.04 & 24.3 & 10.9 & 6.4 & 0.87  \\
\hspace{1cm}... C26 & 20:15:48.8 & +36:31:52.9 & 0.35$\pm$0.03 & 22.1 & \nodata & 2.34 & \nodata  \\
\hspace{1cm}... C27 & 20:15:29.9 & +36:35:31.2 & 0.32$\pm$0.03 & 11.8 & \nodata & 2.34 & \nodata  \\
\hspace{1cm}... C28 & 20:15:52.8 & +36:40:26.0 & 0.55$\pm$0.05 & 13.1 & \nodata & 0.2 & \nodata  \\
\hspace{1cm}... C29 & 20:15:42.4 & +36:41:00.1 & 0.32$\pm$0.03 & 13.3 & \nodata & 0.92 & \nodata  \\
\hspace{1cm}... C30 & 20:15:38.9 & +36:40:24.2 & 0.6$\pm$0.05 & 13.2 & \nodata & 0.61 & \nodata  \\
\hspace{1cm}... C31 & 20:15:48.1 & +36:40:40.7 & 0.52$\pm$0.05 & 8.7 & \nodata & 2.54 & \nodata  \\
\hspace{1cm}... C32 & 20:16:09.7 & +36:39:02.6 & 0.39$\pm$0.04 & 22.4 & \nodata & 0.51 & \nodata  \\
\hspace{1cm}... C33 & 20:17:54.3 & +36:41:40.4 & 0.38$\pm$0.03 & 9.0 & \nodata & 0.61 & \nodata  \\
\hspace{1cm}... C34 & 20:17:58.2 & +36:41:54.7 & 0.60$\pm$0.06 & 11.2 & \nodata & 0.71 & \nodata  \\
\hspace{1cm}... C35 & 20:17:59.9 & +36:47:02.7 & 0.46$\pm$0.04 & 18.6 & \nodata & 0.71 & \nodata  \\

{\bf S148-S149}&&&&&\\
\hspace{1cm}... C1 & 22:56:44.9 & +58:30:06.24 & $1.07\pm0.11$ & 26.7 & 7.4 & -55.05 & -55.67  \\
\hspace{1cm}... C2 & 22:56:43.9 & +58:28:14.88 & 1.03$\pm$0.11 & 22.3 & 9.8 & -54.97 & -54.26  \\
\hspace{1cm}... C3 & 22:56:18.8 & +58:33:14.99 & 0.87$\pm$0.09 & 27.0 & 9.2 & -55.08 & -55.09  \\
\hspace{1cm}... C4 & 22:56:02.3 & +58:33:13.90 & 0.94$\pm$0.1 & 25.8 & 10.6 & -55 & -55.46  \\
\hspace{1cm}... C5 & 22:56:17.2 & +58:31:02.99 & 0.94$\pm$0.1 & 35.0 & 14.6 & -52.58 & -52.14  \\
\hspace{1cm}... C6 & 22:56:26.1 & +58:30:55.00 & 1.06$\pm$0.11 & 24.8 & 9.1 & -54.21 & -54.05  \\
\hspace{1cm}... C7 & 22:56:19.1 & +58:29:48.99 & 0.77$\pm$0.07 & 21.8 & 5.4 & -51.67 & -51.48  \\
\hspace{1cm}... C8 & 22:56:42.3 & +58:31:51.89 & 0.90$\pm$0.1 & 18.8 & 2.5 & -54.17 & -54.33  \\
\hspace{1cm}... C9 & 22:56:21.6 & +58:32:32.49 & 0.84$\pm$0.09 & 18.2 & 8.3 & -53.94 & -51.46  \\
\hspace{1cm}... C10 & 22:56:13.7 & +58:33:15.98 & 0.86$\pm$0.09 & 23.8 & 2.6 & -54.53 & -54.66  \\
\hspace{1cm}... C11 & 22:56:03.9 & +58:31:34.90 & 0.74$\pm$0.08 & 20.4 & 2.8 & -53.7 & -53.35  \\
\hspace{1cm}... C12 & 22:55:56.8 & +58:29:11.84 & 0.75$\pm$0.08 & 20.4 & 5.8 & -54.81 & -54.8  \\
\hspace{1cm}... C13 & 22:56:13.5 & +58:30:05.48 & 0.55$\pm$0.06 & 16.0 & \nodata & -53.1 & \nodata  \\

{\bf S152}&&&&&\\
\hspace{1cm}... C1 & 22:58:56.1 & +58:46:29.27 & $0.32\pm0.03$ & 20.3 & 15.9 & -50.68 & -50.87  \\
\hspace{1cm}... C2-a & 22:58:44.3 & +58:46:51.50 & 0.33$\pm$0.03 & 35.7 & \nodata & -50.72 & \nodata  \\
\hspace{1cm}... C2-b & 22:58:40.6 & +58:46:31.67 & 0.32$\pm$0.03 & 43.9 & 20.9 & -49.85 & -49.84  \\
\hspace{1cm}... C2-c & 22:58:38.1 & +58:46:44.50 & 0.33$\pm$0.03 & 35.3 & \nodata & -49.77 & \nodata  \\
\hspace{1cm}... C3 & 22:58:39.7 & +58:48:15.50 & 0.37$\pm$0.03 & 30.0 & 20.5 & -49.25 & -49.79  \\
\hspace{1cm}... C4 & 22:58:47.7 & +58:45:19.32 & 0.42$\pm$0.04 & 27.4 & 17.0 & -50.84 & -50.74  \\
\hspace{1cm}... C5 & 22:58:59.5 & +58:45:32.08 & 0.31$\pm$0.03 & 21.0 & 12.1 & -50.68 & -50.71  \\
\hspace{1cm}... C6 & 22:58:59.7 & +58:48:29.41 & 0.34$\pm$0.03 & 20.8 & 9.4 & -50.92 & -50.95  \\
\hspace{1cm}... C7 & 22:59:04.0 & +58:49:54.53 & 0.37$\pm$0.03 & 18.9 & 9.0 & -50.84 & -50.58  \\
\hspace{1cm}... C8 & 22:58:31.8 & +58:45:12.32 & 0.25$\pm$0.02 & 22.2 & 7.5 & -51.79 & -49.86  \\

{\bf S175A}&&&&&\\
\hspace{1cm}... C1 & 0:27:29.6 & +64:43:41.30 & 0.13$\pm$0.024 & 28.4 & 12.2 & -49.39 & -49.14  \\
\hspace{1cm}... C2 & 0:27:27.5 & +64:44:45.21 & 0.13$\pm$0.024 & 23.0 & 5.8 & -48.98 & -49.03  \\
\hspace{1cm}... C3 & 0:27:22.0 & +64:45:34.30 & 0.17$\pm$0.033 & 17.8 & 7.4 & -49.39 & -49.03  \\
\hspace{1cm}... C4 & 0:27:19.8 & +64:44:52.34 & 0.12$\pm$0.023 & 27.2 & 13.7 & -49.39 & -49.14  \\
\hspace{1cm}... C5 & 0:27:04.6 & +64:43:49.51 & 0.16$\pm$0.031 & 42.2 & 25.1 & -49.39 & -49.35  \\
\hspace{1cm}... C6 & 0:27:02.3 & +64:43:28.53 & 0.16$\pm$0.031 & 41.2 & 20.9 & -49.79 & -49.67  \\
\hspace{1cm}... C7 & 0:27:28.5 & +64:43:42.19 & 0.15$\pm$0.028 & 29.9 & 8.3 & -49.39 & -49.56  \\
\hspace{1cm}... C8 & 0:27:15.4 & +64:42:46.60 & 0.22$\pm$0.042 & 33.7 & 14.3 & 50 & -49.99  \\
\hspace{1cm}... C9 & 0:26:55.7 & +64:43:35.58 & 0.17$\pm$0.033 & 30.7 & 16.8 & -50.2 & -49.67  \\
\hspace{1cm}... C10 & 0:26:51.4 & +64:43:35.58 & 0.17$\pm$0.033 & 21.7 & 10.7 & -50.81 & -50.52  \\
\hspace{1cm}... C11 & 0:26:41.5 & +64:42:32.59 & 0.19$\pm$0.037 & 15.7 & 2.7 & -50.81 & -50.62  \\
\hspace{1cm}... C12 & 0:26:41.5 & +64:43.35.58 & 0.16$\pm$0.032 & 15.8 & 3.3 & -50.01 & -50.84  \\
\hspace{1cm}... C13 & 0:26:37.2 & +64:44:38.57 & 0.16$\pm$0.030 & 13.9 & 5.2 & -50.01 & -50.94  \\

{\bf S175B}&&&&&\\
\hspace{1cm}... C1 & 0:26:05.4 & +64:54:20.87 & 0.18$\pm$0.04 & 17.5 & 7.1 & -52.03 & -51.12  \\
\hspace{1cm}... C2 & 0:26:19.7 & +64:53:32.71 & 0.14$\pm$0.03 & 20.0 & 3.8 & -51.62 & -51.12  \\
\hspace{1cm}... C3 & 0:26:10.9 & +64:53:03.86 & 0.11$\pm$0.02 & 13.3 & 5.1 & -52.23 & -51.99  \\
\hspace{1cm}... C4 & 0:25:55.5 & +64:51:18.84 & 0.26$\pm$0.05 & 14.3 & 6.2 & -52.03 & -51.22  \\
\hspace{1cm}... C5 & 0:26:17.5 & +64:51:18.84 & 0.24$\pm$0.05 & 24.4 & 5.8 & -49.8 & -49.75  \\
\hspace{1cm}... C6 & 0:26:23.0 & +64:52:49.81 & 0.14$\pm$0.03 & 15.1 & 5.0 & -49.39 & -49.56  \\
\hspace{1cm}... C7 & 0:26:34.0 & +64:52:49.71 & 0.25$\pm$0.05 & 13.5 & 6.5 & -50.61 & -49.57  \\
\hspace{1cm}... C8 & 0:26:31.8 & +64:53:24.73 & 0.24$\pm$0.05 & 13.8 & 8.9 & -50.81 & -49.49  \\
\hspace{1cm}... C9 & 0:26:32.9 & +64:54:27.72 & 0.26$\pm$0.05 & 15.5 & 9.3 & -50.61 & -49.49  \\
\hspace{1cm}... C10 & 0:26:28.5 & +64:54:06.77 & 0.21$\pm$0.04 & 18.6 & 8.3 & -48.78 & -49.22  \\
\hspace{1cm}... C11 & 0:26:40.6 & +64:54:55.63 & 0.17$\pm$0.03 & 14.6 & 9.2 & -48.78 & -49.3  \\
\hspace{1cm}... C12 & 0:26:50.4 & +64:51:18.64 & 0.32$\pm$0.06 & 15.0 & \nodata & -48.78 & \nodata  \\

{\bf S192-S194}&&&&&\\
\hspace{1cm}... C1 & 2:47:48.1 & +61:55:41.67 & 0.35$\pm$0.06 & 16.9 & 7.9 & -44.59 & -44.67  \\
\hspace{1cm}... C2 & 2:47:48.0 & +61:54:45.68 & 0.23$\pm$0.04 & 10.8 & 2.4 & -45.4 & -45.52  \\
\hspace{1cm}... C3 & 2:47:35.2 & +61:56:59.35 & 0.33$\pm$0.06 & 20.8 & 3.6 & -46.01 & -46.05  \\
\hspace{1cm}... C4 & 2:47:35.3 & +61:57:55.35 & 0.24$\pm$0.04 & 16.3 & 4.5 & -46.01 & -46.16  \\
\hspace{1cm}... C5 & 2:47:27.3 & +61:57:27.73 & 0.35$\pm$0.06 & 18.5 & 7.7 & -47.03 & -47.54  \\
\hspace{1cm}... C6 & 2:47:26.3 & +61:56:38.78 & 0.53$\pm$0.09 & 22.9 & 11.1 & -46.83 & -46.8  \\
\hspace{1cm}... C7 & 2:47:25.4 & +61:59:05.82 & 0.33$\pm$0.06 & 12.9 & 3.4 & -46.62 & -46.8  \\
\hspace{1cm}... C8 & 2:47:12.4 & +61:56:25.37 & 0.45$\pm$0.08 & 15.9 & 3.0 & -47.44 & -47.97  \\
\hspace{1cm}... C9 & 2:47:31.1 & +61:54:46.55 & 0.35$\pm$0.06 & 10.8 & 2.1 & -45.4 & -48.71  \\
\hspace{1cm}... C10 & 2:47:15.5 & +61:59:27.25 & 0.36$\pm$0.06 & 8.1 & \nodata & -49.47 & \nodata  \\

{\bf S196}&&&&&\\
\hspace{1cm}... C1 & 2:52:06.7 & +62:09:58.04 & 0.56$\pm$0.12 & 15.1 & \nodata & -42.56 & \nodata  \\
\hspace{1cm}... C2 & 2:51:59.7 & +62:09:30.46 & 57$\pm$0.12 & 16.5 & \nodata & -43.98 & \nodata  \\
\hspace{1cm}... C3 & 2:51:38.7 & +62:09:31.56 & 29$\pm$0.06 & 15.1 & 1.3 & -43.78 & -43.72  \\
\hspace{1cm}... C4 & 2:51:14.8 & +62:11:45.58 & 0.53$\pm$0.11 & 16.5 & 4.6 & -44.59 & -44.57  \\
\hspace{1cm}... C5 & 2:51:22.9 & +62:13:44.27 & 0.45$\pm$0.1 & 7.0 & 1.0 & -45 & -44.67  \\
\hspace{1cm}... C6 & 2:51:28.0 & +62:13:14.05 & 0.45$\pm$0.10 & 7.3 & 0.6 & -45.41 & -45.31  \\
\hspace{1cm}... C7 & 2:51:36.8 & +62:10:27.70 & 0.51$\pm$0.10 & 13.1 & 4.4 & -47.23 & -47.54  \\
\hspace{1cm}... C8 & 2:51:44.8 & +62:10:34.27 & 0.51$\pm$0.11 & 12.1 & 2.6 & -46.62 & -46.59  \\

{\bf S212}&&&&&\\
\hspace{1cm}... C1 & 4:40:34.2 & +50:27:05.98 & 0.72$\pm$0.07 & 31.3 & 5.1 & -33.05 & -35.11  \\
\hspace{1cm}... C2 & 4:40:38.7 & +50:27:05.73 & 0.9$\pm$0.09 & 29.3 & 10.2 & -33.46 & -33.51  \\
\hspace{1cm}... C3 & 4:40:42.4 & +50:27:54.51 & 1.11$\pm$0.11 & 33.0 & 10.8 & -32.85 & -33.09  \\
\hspace{1cm}... C4 & 4:40:47.5 & +50:27:40.21 & 1.0$\pm$0.1 & 23.2 & 6.0 & -35.08 & -35.42  \\
\hspace{1cm}... C5 & 4:40:50.4 & +50:27:40.02 & 0.94$\pm$0.09 & 22.1 & 4.9 & -35.29 & -35.21  \\
\hspace{1cm}... C6 & 4:40:54.0 & +50:27:04.75 & 0.97$\pm$0.1 & 20.6 & 2.9 & -35.69 & -35.74  \\
\hspace{1cm}... C7 & 4:41:01.4 & +50:28:14.24 & 1.25$\pm$0.12 & 17.7 & 5.2 & -35.49 & -36.49  \\
\hspace{1cm}... C8 & 4:41:04.3 & +50:27:04.03 & 1.35$\pm$0.13 & 15.6 & 2.9 & -35.69 & -36.49  \\
\hspace{1cm}... C9 & 4:40:48.3 & +50:28:59.76 & 0.95$\pm$0.09 & 10.5 & 1.9 & -39.15 & -38.82  \\
\hspace{1cm}... C10 & 4:40:45.2 & +50:26:30.33 & 0.86$\pm$0.09 & 12.6 & 2.0 & -34.07 & -34.26  \\
\hspace{1cm}... C11 & 4:40:43.1 & +50:27:26.46 & 0.85$\pm$0.08 & 16.0 & 3.2 & -34.68 & -35  \\
\hspace{1cm}... C12 & 4:40:39.4 & +50:26:23.69 & 0.73$\pm$0.07 & 15.7 & 2.5 & -33.05 & -35  \\

{\bf S288}&&&&&\\
\hspace{1cm}... C1 & 7:08:39.2 & -04:19:21.80 & 0.55$\pm$0.22 & 24.9 & \nodata & 58.38 & \nodata  \\
\hspace{1cm}... C2 & 7:08:37.4 & -04:18:53.80 & 0.54$\pm$0.21 & 28.0 & \nodata & 56.16 & \nodata  \\
\hspace{1cm}... C3 & 7:08:34.6 & -04:17:15.80 & 0.61$\pm$0.25 & 9.9 & \nodata & 56.56 & \nodata  \\
\hspace{1cm}... C4 & 7:08:42.5 & -4:18:46.80 & 0.35$\pm$0.14 & 11.3 & \nodata & 56.95 & \nodata  \\

{\bf S305}&&&&&\\
\hspace{1cm}... C1 & 7:30:04.7 & -18:30:25.27 & 0.75$\pm$0.20 & 25.8 & 9.3 & 42.64 & 43.29  \\
\hspace{1cm}... C2 & 7:30:12.6 & -18:32:03.82 & 0.86$\pm$0.23 & 33.0 & 9.6 & 43.66 & 42.87  \\
\hspace{1cm}... C3 & 7:29:59.3 & -18:30:52.89 & 0.64$\pm$0.17 & 32.1 & 8.6 & 43.86 & 43.93  \\
\hspace{1cm}... C4 & 7:30:16.6 & -18:31:15.08 & 0.61$\pm$0.17 & 26.1 & 11.0 & 44.88 & 44.14  \\
\hspace{1cm}... C5 & 7:29:59.8 & -18:32:37.92 & 0.51$\pm$0.14 & 24.7 & 9.2 & 44.06 & 42.77  \\
\hspace{1cm}... C6 & 7:29:59.8 & -18:32:37.92 & 0.63$\pm$0.17 & 20.2 & 2.8 & 44.67 & 44.46  \\
\hspace{1cm}... C7 & 7:30:00.3 & -18:31:48.96 & 0.62$\pm$0.17 & 59.2 & 14.5 & 46.91 & 46.69  \\
\hspace{1cm}... C8 & 7:30:06.2 & -18:32:59.38 & 0.57$\pm$0.15 & 23.6 & 3.0 & 46.3 & 46.69  \\
\hspace{1cm}... C9 & 7:30:00.3 & -18:33:33.96 & 0.64$\pm$0.17 & 35.1 & 16.4 & 47.11 & 46.8  \\
\hspace{1cm}... C10 & 7:30:02.7 & -18:33:34.13 & 0.57$\pm$0.15 & 16.8 & 2.7 & 47.72 & 48.39  \\
\hspace{1cm}... C11 & 7:30:14.0 & -18:34:37.91 & 0.70$\pm$0.19 & 12.0 & 2.3 & 47.32 & 44.67  \\
\hspace{1cm}... C12 & 7:30:17.0 & -18:33:56.11 & 0.85$\pm$0.23 & 16.3 & \nodata & 44.07 & \nodata  \\

{\bf S307}&&&&&\\
\hspace{1cm}... C1 & 7:35:33.3 & -18:44:59.00 & 0.25$\pm$0.06 & 25.2 & 10.5 & 44.53 & 44.35  \\
\hspace{1cm}... C2-a & 7:35:33.3 & -18:45:20.50 & 0.28$\pm$0.06 & 28.1 & 9.1 & 45.26 & 44.68  \\
\hspace{1cm}... C2-b & 7:35:33.2 & -18:45:34.50 & 0.26$\pm$0.06 & 32.5 & 6.9 & 45.842 & 46.42  \\
\hspace{1cm}... C3 & 7:35:36.3 & -18:46:25.83 & 0.43$\pm$0.1 & 38.0 & 11.2 & 46.6 & 46.55  \\
\hspace{1cm}... C4 & 7:35:38.7 & -18:48:57.50 & 0.47$\pm$0.12 & 44.5 & 19.5 & 47.04 & 46.72  \\
\hspace{1cm}... C5 & 7:35:31.8 & -18:46:23.50 & 0.32$\pm$0.07 & 28.2 & 5.0 & 46.78 & 46.88  \\
\hspace{1cm}... C6-a & 7:35:41.7 & -18:46:16.49 & 0.35$\pm$0.08 & 19.1 & 1.4 & 46.01 & 46.18  \\
\hspace{1cm}... C6-b & 7:35:42.5 & -18:46:30.49 & 0.34$\pm$0.08 & 12.7 & 1.7 & 48.22 & 48.25  \\
\hspace{1cm}... C7 & 7:35:37.4 & -18:44:54.83 & 0.23$\pm$0.05 & 15.0 & 3.7 & 45.8 & 47.79  \\
\hspace{1cm}... C8 & 7:35:43.1 & -18:48:35.32 & 0.37$\pm$0.09 & 26.9 & 8.4 & 47.11 & 47.21  \\
\hspace{1cm}... C9 & 7:35:43.7 & -18:43:55.32 & 0.23$\pm$0.05 & 12.1 & 2.4 & 47.06 & 47.13  \\
\hspace{1cm}... C10 & 7:35:41.0 & -18:44:59.49 & 0.32$\pm$0.07 & 14.3 & 3.6 & 47.29 & 47.34  \\
\enddata
\end{deluxetable}

\clearpage

\begin{deluxetable}{ccccccclcc}
\tabletypesize{\scriptsize}
\tablecaption{Physical parameters calculated for all clumps
\label{tbl3}}
\tablewidth{0pt}
\tablehead{
\colhead{Clump}&\colhead{$T_{ex}$}&\colhead{$ \Delta V_{12}$} & \colhead{$ \Delta V_{13}$} &
\colhead{$N(H)$} &\colhead{$n_{int}(H_2)$} & \colhead{M$_{int}$}&
\colhead{$\tau_{12}$} & \colhead{$\tau_{13}$ }\\
\colhead{No.}&\colhead{(K)}&\colhead{(km s$^{-1}$)}&\colhead{(km s$^{-1}$)}&
\colhead{$(\times10^{20}$cm$^{-2})$}&\colhead{(cm$^{-3}$)}& \colhead{(M$_{\odot}$)}}
\startdata
{\bf S104}&&&&&\\
\hspace{1cm}...C1 & 18.46 & 2.37 & 1.02 & 18.13 & 34.5 & 53$\pm$9 & 17 & 0.27\\
\hspace{1cm}...C2 & 15.631 & 2.81 & 3.05 & 31.77 & 11.9 & 13$\pm$2 & 13 & 0.21\\
\hspace{1cm}...C3 & 25.844 & 3.56 & \nodata & \nodata & 34.3 & 45$\pm$8 & \nodata & \nodata\\
\hspace{1cm}...C4 & 50.193 & 2.16 & 1.91 & 427.26 & 41.7 & 92$\pm$15.9 & 33 & 0.54\\
\hspace{1cm}...C5 & 17.77 & 1.78 & \nodata & \nodata & 14.4 & 8$\pm$1.3 & \nodata & \nodata\\
\hspace{1cm}...C6 & 16.483 & 3.52 & 1.55 & 31.81 & 16.3 & 10$\pm$1.6 & 24 & 0.38\\
\hspace{1cm}...C7 & 34.219 & 2.62 & 0.79 & 42.39 & 33.5 & 59$\pm$10 & 16 & 0.27\\
\hspace{1cm}...C8 & 40.132 & 2.47 & 0.71 & 131.53 & 42.1 & 85$\pm$15 & 42 & 0.68\\
\hspace{1cm}...C9 & 27.714 & 2.71 & 2.15 & 190.27 & 35.9 & 51$\pm$9 & 40 & 0.64\\
\hspace{1cm}...C10 & 39.111 & 3.55 & 2.07 & 243.36 & 44.2 & 65$\pm$11 & 28 & 0.45\\
\hspace{1cm}...C11 & 32.405 & 2.49 & 1.48 & 100.94 & 36.3 & 52$\pm$9 & 23 & 0.37\\
\hspace{1cm}...C12 & 31.932 & 2.12 & 1.87 & 210.72 & 35.4 & 65$\pm$11 & 39 & 0.63\\
\hspace{1cm}...C13 & 37.462 & 2.57 & 1.19 & 146.19 & 46.1 & 51$\pm$9 & 32 & 0.51\\
\hspace{1cm}...C14 & 37.352 & 2.69 & 1.87 & 102.59 & 41.6 & 46$\pm$8 & 14 & 0.23\\
\hspace{1cm}...C15 & 27.957 & 2.05 & 1.26 & 78.92 & 21.2 & 44$\pm$8 & 28 & 0.45\\
\hspace{1cm}...C16 & 28.711 & 2.47 & 1.97 & 159.49 & 28.0 & 34$\pm$6 & 34 & 0.55\\
\hspace{1cm}...C17 & 31.784 & 2.13 & 1.63 & 51.45 & 15.8 & 19$\pm$3 & 11 & 0.18\\
\hspace{1cm}...C18 & 20.363 & 2.99 & 1.67 & 21.06 & 13.1 & 23$\pm$4 & 10 & 0.16\\
\hspace{1cm}...C19 & 29.326 & 2.85 & 2.02 & 73.27 & 24.2 & 36$\pm$6 & 15 & 0.24\\
\hspace{1cm}...C20 & 17.82 & 1.87 & 1.31 & 10.42 & 9.5 & 8$\pm$1.4 & 8 & 0.13\\
\hspace{1cm}...C21 & 27.45 & 2.31 & 1.89 & 51.04 & 39.6 & 17$\pm$2.9 & 12 & 0.20\\
\hspace{1cm}...C22 & 26.855 & 2.77 & \nodata & \nodata & 26.2 & 19$\pm$3 & \nodata & \nodata\\
\hspace{1cm}...C23 & 19.414 & 2.4 & 1.64 & 33.31 & 19.6 & 9$\pm$1.6 & 18 & 0.28\\
\hspace{1cm}...C24 & 23.031 & 1.29 & 1.06 & 25.51 & 9.0 & 7$\pm$1.3 & 15 & 0.25\\
\hspace{1cm}...C25 & 29.66 & 2.2 & 2.22 & 204.5 & 14.7 & 31$\pm$5 & 37 & 0.59\\
\hspace{1cm}...C26 & 27.419 & 2.17 & \nodata & \nodata & 32.3 & 32$\pm$6 & \nodata & \nodata\\
\hspace{1cm}...C27 & 16.936 & 2.34 & \nodata & \nodata & 18.9 & 15$\pm$2.6 & \nodata & \nodata\\
\hspace{1cm}...C28 & 18.289 & 3.16 & \nodata & \nodata & 15.0 & 59$\pm$10 & \nodata & \nodata\\
\hspace{1cm}...C29 & 18.468 & 1.53 & \nodata & \nodata & 14.3 & 11$\pm$2 & \nodata & \nodata\\
\hspace{1cm}...C30 & 18.421 & 1.87 & \nodata & \nodata & 9.9 & 51$\pm$9 & \nodata & \nodata\\
\hspace{1cm}...C31 & 13.665 & 2.01 & \nodata & \nodata & 14.2 & 48$\pm$8 & \nodata & \nodata\\
\hspace{1cm}...C32 & 27.729 & 2.26 & \nodata & \nodata & 28.1 & 40$\pm$7 & \nodata & \nodata\\
\hspace{1cm}...C33 & 14.057 & 1.98 & \nodata & \nodata & 9.7 & 13$\pm$2.2 & \nodata & \nodata\\
\hspace{1cm}...C34 & 16.305 & 1.88 & \nodata & \nodata & 6.7 & 35$\pm$6 & \nodata & \nodata\\
\hspace{1cm}...C35 & 23.883 & 2.8 & \nodata & \nodata & 22.0 & 51$\pm$9 & \nodata & \nodata\\

{\bf S148-S149}&&&&&\\
\hspace{1cm}...C1 & 32.079 & 4.21 & 2.62 & 152.34 & 47.9 & 1400$\pm$290 & 20 & 0.32\\
\hspace{1cm}...C2 & 27.708 & 6.75 & 2.79 & 218.55 & 48.8 & 1280$\pm$260 & 35 & 0.57\\
\hspace{1cm}...C3 & 32.395 & 1.86 & 0.92 & 69.01 & 27.3 & 435$\pm$89 & 25 & 0.41\\
\hspace{1cm}...C4 & 31.238 & 1.67 & 1.36 & 122.4 & 20.9 & 412$\pm$84 & 32 & 0.52\\
\hspace{1cm}...C5 & 40.48 & 3.27 & 2.42 & 361.84 & 48.1 & 960$\pm$196 & 33 & 0.54\\
\hspace{1cm}...C6 & 30.229 & 3.83 & 1.04 & 76.15 & 31.3 & 900$\pm$184 & 28 & 0.45\\
\hspace{1cm}...C7 & 27.16 & 3.71 & 1.57 & 58.08 & 35.7 & 900$\pm$77 & 17 & 0.28\\
\hspace{1cm}...C8 & 24.099 & 2.19 & 1.65 & 25.16 & 18.5 & 386$\pm$67 & 9 & 0.14\\
\hspace{1cm}...C9 & 23.551 & 2.19 & 1 & 60.59 & 10.3 & 328$\pm$30 & 37 & 0.60\\
\hspace{1cm}...C10 & 29.216 & 2.15 & 1.65 & 28.16 & 29.0 & 440$\pm$90 & 7 & 0.11\\
\hspace{1cm}...C11 & 25.703 & 2.38 & 1.53 & 26.79 & 21.4 & 211$\pm$43 & 9 & 0.15\\
\hspace{1cm}...C12 & 25.727 & 1.39 & 0.9 & 34.43 & 11.6 & 118$\pm$24 & 21 & 0.34\\
\hspace{1cm}...C13 & 21.24 & 2.27 & \nodata & \nodata & 21.3 & 191$\pm$53 & \nodata & \nodata\\

{\bf S152}&&&&&\\
\hspace{1cm}...C1 & 25.647 & 5.28 & 1.37 & 508.99 & 156.0 & 117$\pm$18 & 93 & 1.51\\
\hspace{1cm}...C2-a & 41.211 & 4.66 & \nodata & \nodata & 207.0 & 176$\pm$28 & \nodata & \nodata\\
\hspace{1cm}...C2-b & 49.415 & 4.35 & 1.75 & 456.5 & 233.0 & 175$\pm$28 & 40 & 0.64\\
\hspace{1cm}...C2-c & 40.782 & 4.87 & \nodata & \nodata & 213.0 & 181$\pm$29 & \nodata & \nodata\\
\hspace{1cm}...C3 & 35.468 & 4.76 & 1.69 & 416.42 & 151.0 & 189$\pm$30 & 70 & 1.13\\
\hspace{1cm}...C4 & 32.78 & 7.42 & 3.15 & 566.29 & 165.0 & 295$\pm$47 & 60 & 0.96\\
\hspace{1cm}...C5 & 26.311 & 2.43 & 1.59 & 168.75 & 77.0 & 56$\pm$9 & 53 & 0.85\\
\hspace{1cm}...C6 & 26.121 & 3.04 & 2.04 & 148.31 & 84.0 & 82$\pm$13 & 37 & 0.59\\
\hspace{1cm}...C7 & 24.242 & 2.43 & 1.07 & 73.02 & 519.0 & 633$\pm$100 & 39 & 0.63\\
\hspace{1cm}...C8 & 27.516 & 6.04 & 2.52 & 139.02 & 268.0 & 97$\pm$15 & 25 & 0.41\\

{\bf S175A}&&&&&\\
\hspace{1cm}...C1 & 33.828 & 0.97 & 0.56 & 61.83 & 118.9 & 6$\pm$2.0 & 34 & 0.56\\
\hspace{1cm}...C2 & 28.362 & 0.78 & 0.47 & 19.38 & 82.4 & 4$\pm$1.4 & 18 & 0.29\\
\hspace{1cm}...C3 & 23.076 & 1.12 & 0.66 & 34.68 & 49.6 & 6$\pm$2.1 & 33 & 0.54\\
\hspace{1cm}...C4 & 32.626 & 0.92 & 0.56 & 72.12 & 118.7 & 5$\pm$1.6 & 43 & 0.69\\
\hspace{1cm}...C5 & 47.676 & 1.31 & 0.85 & 290.4 & 139.7 & 14$\pm$5 & 56 & 0.90\\
\hspace{1cm}...C6 & 46.698 & 1.61 & 0.95 & 243.38 & 201.2 & 19$\pm$7 & 44 & 0.70\\
\hspace{1cm}...C7 & 35.307 & 0.97 & 0.64 & 44.75 & 113.7 & 9$\pm$3.1 & 20 & 0.33\\
\hspace{1cm}...C8 & 39.128 & 1.12 & 0.74 & 105.94 & 71.4 & 18$\pm$6.4 & 34 & 0.55\\
\hspace{1cm}...C9 & 36.178 & 1.54 & 1.08 & 190.75 & 126.8 & 15$\pm$5.2 & 49 & 0.78\\
\hspace{1cm}...C10 & 27.025 & 1.59 & 0.93 & 82.41 & 77.1 & 9$\pm$3.2 & 42 & 0.68\\
\hspace{1cm}...C11 & 20.901 & 1.25 & 0.87 & 13.28 & 35.8 & 6$\pm$2.1 & 12 & 0.19\\
\hspace{1cm}...C12 & 21.039 & 1.15 & 0.71 & 13.5 & 63.2 & 7$\pm$2.4 & 14 & 0.23\\
\hspace{1cm}...C13 & 19.084 & 1.32 & 0.62 & 20.09 & 70.6 & 6$\pm$2.3 & 29 & 0.47\\

{\bf S175B}&&&&&\\
\hspace{1cm}...C1 & 22.819 & 8.68 & 1.44 & 70.78 & 220.9 & 31$\pm$11 & 32 & 0.51\\
\hspace{1cm}...C2 & 25.341 & 4.1 & 1.4 & 34.42 & 183.0 & 12$\pm$4.1 & 13 & 0.21\\
\hspace{1cm}...C3 & 18.466 & 4.87 & 1.57 & 49.57 & 178.1 & 6$\pm$2.2 & 30 & 0.48\\
\hspace{1cm}...C4 & 19.52 & 2 & 1.26 & 51.17 & 38.6 & 16$\pm$5.6 & 35 & 0.56\\
\hspace{1cm}...C5 & 29.769 & 1.04 & 0.75 & 31.58 & 48.7 & 17$\pm$5.8 & 17 & 0.27\\
\hspace{1cm}...C6 & 20.339 & 2.6 & 1.09 & 33.82 & 139.0 & 9$\pm$3.1 & 25 & 0.40\\
\hspace{1cm}...C7 & 18.71 & 2.77 & 1.32 & 57.33 & 63.0 & 24$\pm$9 & 40 & 0.65\\
\hspace{1cm}...C8 & 19.019 & 3.3 & 1.22 & 85.83 & 86.2 & 29$\pm$10 & 63 & 1.02\\
\hspace{1cm}...C9 & 20.774 & 2.72 & 1.39 & 100.91 & 64.6 & 26$\pm$9 & 56 & 0.90\\
\hspace{1cm}...C10 & 23.955 & 1.94 & 1.17 & 71.86 & 93.4 & 22$\pm$8 & 36 & 0.59\\
\hspace{1cm}...C11 & 19.838 & 2.6 & 1.4 & 101.21 & 93.3 & 10$\pm$3.6 & 60 & 0.97\\
\hspace{1cm}...C12 & 20.274 & 1.13 & \nodata & \nodata & 16.4 & 12$\pm$4.3 & \nodata & \nodata\\

{\bf S192-S194}&&&&&\\
\hspace{1cm}...C1 & 22.204 & 1.49 & 0.82 & 46.41 & 27.0 & 26$\pm$8 & 38 & 0.62\\
\hspace{1cm}...C2 & 15.906 & 1.29 & 0.83 & 10.26 & 23.6 & 7$\pm$2 & 15 & 0.25\\
\hspace{1cm}...C3 & 26.156 & 2.19 & 1.76 & 40.57 & 71.0 & 61$\pm$19 & 12 & 0.19\\
\hspace{1cm}...C4 & 21.597 & 1.04 & 0.75 & 20.89 & 29.9 & 9$\pm$3.0 & 20 & 0.32\\
\hspace{1cm}...C5 & 23.829 & 2.05 & 0.98 & 53.91 & 58.5 & 57$\pm$18 & 33 & 0.53\\
\hspace{1cm}...C6 & 28.272 & 3.45 & 2.13 & 199.34 & 6.1 & 22$\pm$7 & 41 & 0.66\\
\hspace{1cm}...C7 & 18.041 & 1.76 & 1.2 & 23.06 & 22.6 & 19$\pm$6 & 19 & 0.31\\
\hspace{1cm}...C8 & 21.167 & 1.98 & 1.27 & 21.93 & 28.0 & 61$\pm$20 & 13 & 0.21\\
\hspace{1cm}...C9 & 15.906 & 1.28 & 0.63 & 6.64 & 14.1 & 14$\pm$4.5 & 13 & 0.21\\
\hspace{1cm}...C10 & 13.057 & 0.94 & \nodata & \nodata & 6.6 & 8$\pm$2.4 & \nodata & \nodata\\

{\bf S196}&&&&&\\
\hspace{1cm}...C1 & 20.336 & 3.09 & \nodata & \nodata & 42.9 & 179$\pm$73 & \nodata & \nodata\\
\hspace{1cm}...C2 & 21.804 & 2.83 & \nodata & \nodata & 44.6 & 201$\pm$82 & \nodata & \nodata\\
\hspace{1cm}...C3 & 20.336 & 0.83 & 0.71 & 4.96 & 11.2 & 6$\pm$2.6 & 6 & 0.09\\
\hspace{1cm}...C4 & 21.804 & 1.7 & 0.92 & 25.97 & 16.4 & 58$\pm$24 & 20 & 0.32\\
\hspace{1cm}...C5 & 11.839 & 2.05 & 1.15 & 5.12 & 10.1 & 23$\pm$9 & 9 & 0.15\\
\hspace{1cm}...C6 & 12.203 & 1.37 & 1.4 & 3.63 & 5.7 & 13$\pm$5.2 & 5 & 0.08\\
\hspace{1cm}...C7 & 18.271 & 1.6 & 1.11 & 28.71 & 18.0 & 58$\pm$24 & 25 & 0.40\\
\hspace{1cm}...C8 & 17.275 & 1.2 & 0.68 & 9.53 & 13.4 & 41$\pm$17 & 15 & 0.24\\

{\bf S212}&&&&&\\
\hspace{1cm}...C1 & 36.717 & 3.29 & 1.48 & 60.59 & 53.5 & 479$\pm$90 & 11 & 0.18\\
\hspace{1cm}...C2 & 34.708 & 2.91 & 1.58 & 139.7 & 38.0 & 671$\pm$125 & 26 & 0.42\\
\hspace{1cm}...C3 & 38.483 & 3.81 & 1.8 & 178.81 & 31.0 & 1028$\pm$193 & 24 & 0.39\\
\hspace{1cm}...C4 & 28.605 & 3.75 & 2.15 & 92.13 & 34.8 & 836$\pm$157 & 18 & 0.30\\
\hspace{1cm}...C5 & 27.506 & 1.57 & 1.09 & 36.84 & 16.7 & 335$\pm$63 & 15 & 0.25\\
\hspace{1cm}...C6 & 25.897 & 2.77 & 1.03 & 18.97 & 21.0 & 463$\pm$87 & 9 & 0.15\\
\hspace{1cm}...C7 & 22.956 & 3.08 & 1.37 & 46 & 15.6 & 738$\pm$138 & 21 & 0.35\\
\hspace{1cm}...C8 & 20.845 & 2.31 & 0.74 & 12.3 & 10.6 & 624$\pm$117 & 13 & 0.20\\
\hspace{1cm}...C9 & 15.565 & 2.67 & 2.2 & 21.3 & 10.7 & 223$\pm$42 & 12 & 0.20\\
\hspace{1cm}...C10 & 17.794 & 2 & 1.35 & 13.94 & 15.2 & 232$\pm$44 & 10 & 0.17\\
\hspace{1cm}...C11 & 21.207 & 2.63 & 1.94 & 35.61 & 29.2 & 430$\pm$81 & 14 & 0.22\\
\hspace{1cm}...C12 & 20.965 & 3.04 & 2.25 & 31.97 & 36.1 & 341$\pm$64 & 11 & 0.17\\

{\bf S288}&&&&&\\
\hspace{1cm}...C1 & 30.282 & 3.84 & \nodata & \nodata & 38.4 & 150$\pm$114 & \nodata & \nodata\\
\hspace{1cm}...C2 & 33.4 & 3.44 & \nodata & \nodata & 44.4 & 164$\pm$125 & \nodata & \nodata\\
\hspace{1cm}...C3 & 14.967 & 1.43 & \nodata & \nodata & 8.1 & 45$\pm$34 & \nodata & \nodata\\
\hspace{1cm}...C4 & 16.42 & 1.66 & \nodata & \nodata & 15.9 & 17$\pm$13 & \nodata & \nodata\\

{\bf S305}&&&&&\\
\hspace{1cm}...C1 & 31.248 & 2.66 & 1.95 & 148.43 & 42.0 & 430$\pm$220 & 28 & 0.44\\
\hspace{1cm}...C2 & 38.498 & 3.42 & 1.72 & 148.74 & 49.4 & 760$\pm$390 & 21 & 0.34\\
\hspace{1cm}...C3 & 37.608 & 2.45 & 1.54 & 115.31 & 57.9 & 361$\pm$185 & 19 & 0.31\\
\hspace{1cm}...C4 & 31.497 & 1.93 & 1.04 & 97.79 & 40.0 & 223$\pm$114 & 34 & 0.54\\
\hspace{1cm}...C5 & 30.091 & 1.74 & 1.13 & 83.23 & 32.6 & 189$\pm$97 & 29 & 0.46\\
\hspace{1cm}...C6 & 25.524 & 1.32 & 1.01 & 17.89 & 33.9 & 107$\pm$55 & 9 & 0.15\\
\hspace{1cm}...C7 & 64.765 & 2.4 & 1.43 & 272.49 & 77.5 & 466$\pm$239 & 17 & 0.28\\
\hspace{1cm}...C8 & 29.006 & 2.68 & 2.14 & 43.7 & 37.6 & 169$\pm$87 & 8 & 0.14\\
\hspace{1cm}...C9 & 40.628 & 1.73 & 1.26 & 220.45 & 41.1 & 256$\pm$131 & 39 & 0.62\\
\hspace{1cm}...C10 & 22.026 & 1.76 & 1.18 & 18.58 & 27.8 & 125$\pm$64 & 11 & 0.18\\
\hspace{1cm}...C11 & 17.174 & 3.49 & 1.4 & 16.67 & 40.9 & 344$\pm$176 & 13 & 0.21\\
\hspace{1cm}...C12 & 21.552 & 3.2 & \nodata & \nodata & 27.5 & 407$\pm$208 & \nodata & \nodata\\

{\bf S307}&&&&&\\
\hspace{1cm}...C1 & 30.617 & 2.18 & 1.11 & 97.93 & 101.3 & 38$\pm$16 & 33 & 0.54\\
\hspace{1cm}...C2-a & 33.508 & 2.48 & 1.11 & 126.02 & 71.3 & 37$\pm$16 & 24 & 0.39\\
\hspace{1cm}...C2-b & 38.003 & 2.41 & 2.26 & 132.25 & 104.8 & 44$\pm$19 & 15 & 0.24\\
\hspace{1cm}...C3 & 43.504 & 2.59 & 2.28 & 251.74 & 78.4 & 152$\pm$66 & 21 & 0.35\\
\hspace{1cm}...C4 & 50.001 & 3.35 & 2.16 & 513.66 & 99.7 & 251$\pm$108 & 35 & 0.57\\
\hspace{1cm}...C5 & 33.622 & 2.22 & 1.89 & 71.85 & 70.3 & 56$\pm$24 & 12 & 0.19\\
\hspace{1cm}...C6-a & 24.456 & 1.58 & 1.03 & 8.67 & 29.8 & 31$\pm$14 & 5 & 0.08\\
\hspace{1cm}...C6-b & 17.912 & 1 & 0.7 & 6.24 & 19.3 & 18$\pm$8 & 9 & 0.14\\
\hspace{1cm}...C7 & 20.192 & 1.62 & 0.6 & 12.99 & 54.2 & 16$\pm$7 & 18 & 0.28\\
\hspace{1cm}...C8 & 32.338 & 2.63 & 1.38 & 93.76 & 100.0 & 124$\pm$54 & 23 & 0.37\\
\hspace{1cm}...C9 & 17.279 & 1.07 & 0.68 & 8.59 & 24.8 & 7$\pm$3.2 & 14 & 0.22\\
\hspace{1cm}...C10 & 19.474 & 1.32 & 1 & 20.46 & 25.0 & 19$\pm$8 & 18 & 0.29\\
\enddata
\end{deluxetable}

\begin{deluxetable}{lllcccc}
\tabletypesize{\scriptsize}
\tablecaption{The Larson power law index  for Type I regions \label{slopes}}
\tablewidth{0pt}
\tablehead{
\colhead{Source}& \colhead{$\alpha[\Delta V_{12}]$} & \colhead{$\alpha[\Delta V_{13}]$} &
}
\startdata

S148-S149&3.0 $\pm$ 0.5 & 2.5 $\pm$ 0.7 &\\
S175A& 1.3 $\pm$ 0.3 & 1.4 $\pm$ 0.3 & \\
S192-194&1.5 $\pm$  0.2 & 1.5  $\pm$ 0.2  & \\
S196&1.9 $\pm$  0.4 &  1.2  $\pm$0.2   & \\
S305&1.9 $\pm$ 0.2  & 1.6 $\pm$ 0.2  & \\
S307&1.4 $\pm$ 0.2 & 1.9 $\pm$ 0.2  & \\
\enddata

\end{deluxetable}

\end{document}